# What should I say? *Interacting with AI and Natural Language Interfaces*

Mark Adkins
College of Computing
Georgia Institute of Technology
Atlanta, GA, United States
madkins@gatech.edu

*Abstract*—As Artificial Intelligence (AI) technology becomes more and more prevalent, it becomes increasingly important to explore how we as humans interact with AI. The Human-AI Interaction (HAI) sub-field has emerged from the Human-Computer Interaction (HCI) field and aims to examine this very notion. Many interaction patterns have been implemented without fully understanding the changes in required cognition as well as the cognitive science implications of using these alternative interfaces that aim to be more human-like in nature. Prior research suggests that theory of mind representations are crucial to successful and effortless communication, however very little is understood when it comes to how theory of mind representations are established when interacting with AI.

## I. Introduction

With the increase in popularity, capability, and usage of artificial intelligence (AI) it has become more and more critical that we understand the ways we use and interact with it. In general, the intent behind natural language and voice interfaces is to provide a method of communicating with technology that is either 1) hands free, or 2) allows us to use our preferred phrasing and sentence structure. The theory of mind concept refers to our ability to create a mental model of who (or what) we are interacting with. This mental model allows us to build expectations and establish a notable level of comfort when interacting with a given entity [1], [2], [6], [7], [8].

A strong theory of mind allows us to communicate in a way that feels natural because we have a general understanding of the person or thing we are interacting with and know approximately what to expect when we communicate externally in a given manner. It is important to note that this is not just regarding verbal communication. This also includes body language and other forms of nonverbal communication [2], [3], [8, Ch. 9], [10]. Other important factors in being able to develop a theory of mind go as far as someone's voice intonation. Intonation and spoken emphasis is important in many languages. For example, in the English language, intonation and spoken emphasis are the most prominent factors in communicating sarcasm. This sentiment is vital to *correctly* interpreting the literal words spoken since the true meaning is often at odds with what was actually said. All of these could be considered "interfaces" of communication. Artificial intelligence has only a single interface to communicate with, making it *theoretically* exceedingly difficult to build a robust theory of mind for. The goal of this research is to prove this, and measure to the degree that this impacts our ability to efficiently and effectively communicate with AI through these "natural language" interfaces. In order to accomplish this, this research aims to answer the following research question:

*Do humans change their natural communication style or behavior when interacting with Artificial Intelligence (AI)?*

The hypothesis tested was *yes*, we as humans in the general population do change our natural communication patterns when interacting with AI.

## II. Experiment Design

To explore and test this hypothesis, a survey was developed to identify patterns in the general population when interacting with AI and natural language interfaces. As part of the latter, both voice and text interfaces were tested.

### A. Survey data type breakdown

This survey consisted of 55 questions with the type of data collection as follows:

- 8 Segmentation questions
- 6 Yes/No boolean questions
- 2 Multi-select questions
- 3 Free response questions
- 11 Frequency scale questions
- 25 Likert scale questions

### B. Themes explored

The goal of the survey instrument was to supplement existing literature to provide new data and insights that would reveal an answer to the aforementioned research question. This survey explored how frequently participants used AI, and what interface(s) they used to engage with AI. The survey was designed to highlight use of voice interfaces (such as voice assistants and customer support screening mechanisms) and text interfaces (such as ChatGPT, Llama, or virtual assistants such as chatbots and customer support virtual agents).

Using these targeted reference points, additional questions were asked to explore frequency of use, confidence of accuracy and understanding, behavior and feelings, pattern adaptation/deviation, and overall sentiment. The control variable in each of these theme explorations was a participant's friend that they knew well. This is based on a literature-backed assumption that a respondent will likely have a robust theory of mind representation for a close friend [7], which provides us a baseline to measure against when asking about targeted areas of interest.

*C. Survey participation*

The original participation goal for this study was 50 adults. Participation was voluntary. The first distribution of the survey received disproportionately higher participation in younger segments such as undergraduate university students. A second round of survey distribution was done to help balance participation segments across age, education level, and professional industries. Only people 18 years of age and older were surveyed for this study, so it will be important to keep in mind that these results are representative of the *adult* general population.

Overall 101 responses were received, just over double the initial target. Participants were permitted to skip questions if they did not feel comfortable providing an answer. Participants were also permitted to end the survey at any time. For this reason, most questions received 92 responses on average with the least answered questions receiving 89 responses.

*D. Populations represented in results*

As previously mentioned, a concerted effort was made to balance segments to maximize reach and mitigate bias. According to segmentation data collected, these responses include 5 education levels, 6 employment situations, every household income bracket included as an option, as well as 11 different professional industries. Participants spanned across 9 countries, with participants from the United States making up the majority of respondents.

*E. Quantifying results*

The 25 Likert scale questions had their responses aggregated and then assigned a numerical weight. The phrasing varied depending on the context of the question, but in general negative responses (such as "Disagree" and "Strongly Disagree") were assigned a -1 and -2 value respectively. Conversely positive responses (such as "Agree" and "Strongly Agree") were assigned a 1 and 2 value respectively. The neutral option (such as "Neither agree nor disagree") was assigned a 0. The sum was taken of all responses for each question and then divided by the total number of responses to get the population average score. By dividing the sum by the total number of responses, the input from neutral responses still affects the population score despite being given no weight.

This quantification was done to observe if the general population leaned positively or negatively when reacting to a given statement.

## III. RESULTS

*A. Theme 1: Communication patterns*

In general, respondents self-reported that they do change their communication patterns when interacting with AI, especially voice assistants. This pattern emerged from looking at multiple questions. There is strong evidence that the overall population generally changes the way they phrase sentences to help an AI understand them better.

1) *I change how I phrase things to AIs to try to help it understand me better:* This statement had a mean population score of 1.02 indicating the overall population falls in the "Somewhat agree" category [Fig. 1].

| Strongly disagree | Somewhat disagree | Neither agree nor disagree | Somewhat agree | Strongly Agree |
|---|---|---|---|---|
| -2 | -1 | 0 | 1 | 2 |
| 0 | 6 | 15 | 39 | 29 |

2) *I think AI has limitations when it comes to understanding the way I speak or write:* This statement had a mean population score of 0.97 indicating the overall population falls in the "Somewhat agree" category [Fig. 2].

| Strongly disagree | Somewhat disagree | Neither agree nor disagree | Somewhat agree | Strongly Agree |
|---|---|---|---|---|
| -2 | -1 | 0 | 1 | 2 |
| 1 | 10 | 15 | 28 | 35 |

3) *I have to think more systematically to communicate with an AI:* This statement had a mean population score of 0.76 indicating the overall population leans toward the "Somewhat agree" category [Fig. 3].

| Strongly disagree | Somewhat disagree | Neither agree nor disagree | Somewhat agree | Strongly Agree |
|---|---|---|---|---|
| -2 | -1 | 0 | 1 | 2 |
| 2 | 11 | 16 | 37 | 23 |

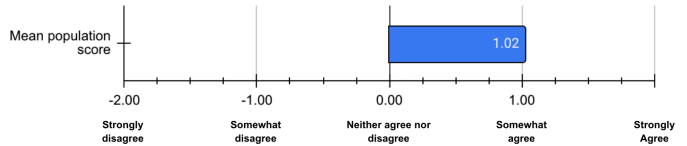

1. The mean population score of the Likert scale question stating "I change how I phrase things to AIs to try to help it understand me better."

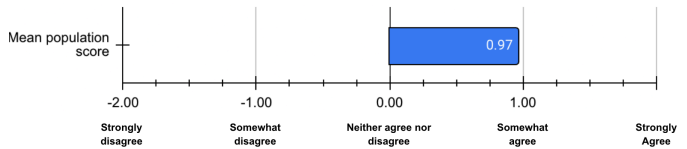

2. The mean population score of the Likert scale question stating "I think AI has limitations when it comes to understanding the way I speak or write."

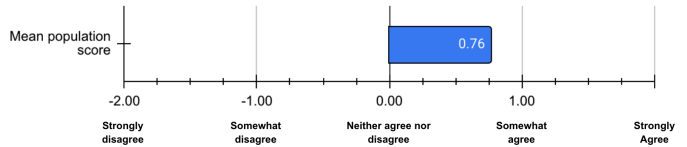

3. The mean population score of the Likert scale question stating "I have to think more systematically to communicate with an AI."

4) *If you use AI-powered chats, what aspects of your sentence structure change when engaging with the AI?:* This was a multi-option follow-up question asked to those who agreed that they changed their sentence structure when using an AI chat. *Length of sentences* was the most common self-reported change, followed by *Use of filler words*. [Fig. 4].

5) *If you use voice interfaces, what aspects of your speech patterns change when using a voice assistant?:* This was a multi-option follow-up question asked to those who agreed that they changed their sentence structure when using a voice assistant. *Enunciation* was the most common self-reported change, followed by a close mix of *Volume*, *Sentence complexity*, and *Pauses and flow*. [Fig. 5].

Participants felt that they deviated from their natural sentence structure more when interacting with voice assistants (mean population score: 0.86) than with AI chat interfaces (mean population score: 0.58). In other questions, there was a notable ambivalence that mildly leaned toward confidence in voice assistants hearing an individual correctly (mean population score 0.16). This is not representative of unconfidence, but is still fairly low overall confidence in voice assistants' ability to hear correctly. This may be one reason for the increased desire to modify one's sentence structure when using a voice assistant.

| Extremely unconfident | Somewhat unconfident | Neither confident nor unconfident | Somewhat confident | Extremely confident |
|---|---|---|---|---|
| -2 | -1 | 0 | 1 | 2 |
| 4 | 25 | 17 | 42 | 3 |

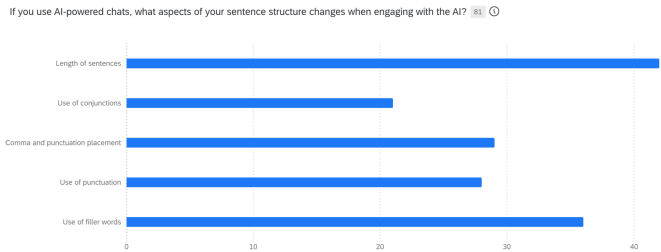

4. The responses to the question asking "If you use voice interfaces, what aspects of your speech patterns change when using a voice assistant?"

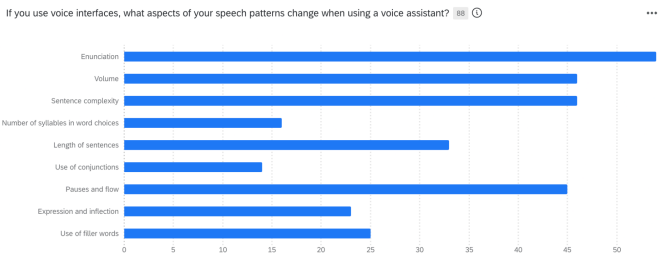

5. The responses to the question asking "If you use AI-powered chats, what aspects of your sentence structure changes when engaging with the AI?"

## B. Theme 2: Differences from interacting with friends

In general, respondents self-reported that they interact with AI in a manner that is different from how they commonly interact with friends. There was also agreement that it does not feel natural. While the scores indicating that interacting with AI does not feel natural are not the strongest consensus in the study, responses do indicate that there is something consistently amiss about it across multiple questions. This theme presented itself in three questions.

The questions measuring against a self-directed control variable (a friend) show consistent agreement that interacting with an AI, regardless of interface or format, is notably different than interacting with a friend. This theme presented itself in an additional four questions.

1) *I phrase things when conversing with an AI in the same way I would with a friend:* This statement had a mean population score of -1.00 indicating the overall population falls in the "Somewhat disagree" category [Fig. 6].

| Strongly disagree | Somewhat disagree | Neither agree nor disagree | Somewhat agree | Strongly Agree |
|---|---|---|---|---|
| -2 | -1 | 0 | 1 | 2 |
| 33 | 34 | 12 | 9 | 1 |

2) *I pay attention to my sentence structure more when conversing with an AI than with a friend:* This statement had a mean population score of 0.83 indicating the overall population falls in the "Somewhat agree" category [Fig. 7].

| Strongly disagree | Somewhat disagree | Neither agree nor disagree | Somewhat agree | Strongly Agree |
|---|---|---|---|---|
| -2 | -1 | 0 | 1 | 2 |
| 2 | 10 | 14 | 38 | 25 |

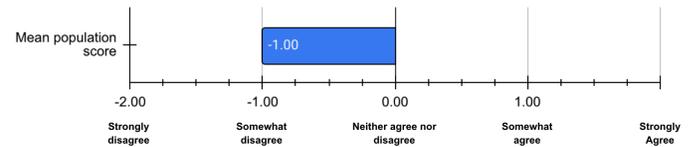

6. The mean population score of the Likert scale question stating "I phrase things when conversing with an AI in the same way I would with a friend."

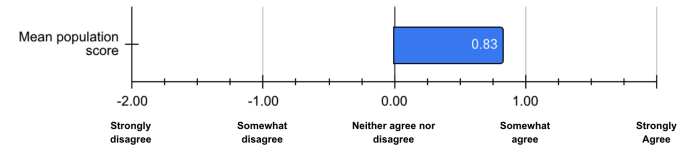

7. The mean population score of the Likert scale question stating "I pay attention to my sentence structure more when conversing with an AI than with a friend."

3) *Interacting with an AI feels natural:* This statement had a mean population score of -0.57 indicating the overall population leans towards the "Somewhat disagree" category [Fig. 8].

| Strongly disagree | Somewhat disagree | Neither agree nor disagree | Somewhat agree | Strongly Agree |
|---|---|---|---|---|
| -2 | -1 | 0 | 1 | 2 |
| 16 | 33 | 26 | 14 | 0 |

4) *Does having a conversation with a voice assistant or AI-powered chat feel comparable to having a conversation with another person?:* This question had a mean population score of -0.65 indicating the overall population leans strongly towards "No" [Fig. 9].

| No | Unsure | Yes |
|---|---|---|
| -1 | 0 | 1 |
| 68 | 5 | 18 |

5) *Does speaking to a voice assistant feel natural to you?:* This question had a mean population score of -0.55 indicating the overall population leans strongly towards "No" [Fig. 10].

| No | Unsure | yes |
|---|---|---|
| -1 | 0 | 1 |
| 63 | 10 | 18 |

Note that the previous theme revealed users tend to alter their sentence structure less when interacting with an AI chat. Despite respondents reporting they change communication patterns more with voice assistants, they also report voice assistants feel *slightly* more natural than AI chat.

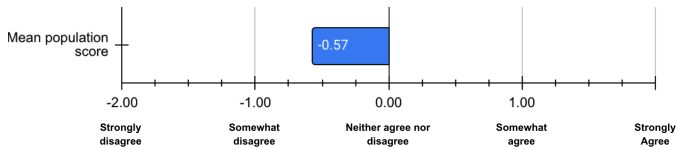

**Interacting with an AI feels natural.**

8. The mean population score of the Likert scale question stating "Interacting with an AI feels natural"

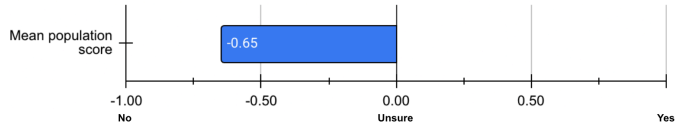

**Does having a conversation with a voice assistant or AI-powered chat feel comparable to having a conversation with another person?**

9. The mean population score of the Yes/No/Unsure question asking "Does having a conversation with a voice assistant or AI-powered chat feel comparable to having a conversation with another person?"

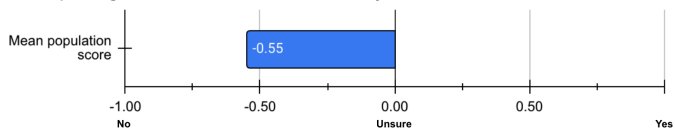

**Does speaking to a voice assistant feel natural to you?**

10. The mean population score of the Yes/No/Unsure question asking "Does speaking to a voice assistant feel natural to you?"

6) *Does having a conversation in an AI-powered chat feel natural to you?:* This statement had a mean population score of -0.49 indicating the overall population leans strongly towards "No" [Fig. 11].

| No | Unsure | Yes |
|---|---|---|
| -1 | 0 | 1 |
| 63 | 10 | 18 |

7) *Do you speak to a voice assistant in the same way you speak to a friend?:* This statement had a mean population score of -1.27 indicating the overall population falls notably beyond the "Probably not" category and leans further towards "Definitely not" [Fig. 12].

| Strongly disagree | Somewhat disagree | Neither agree nor disagree | Somewhat agree | Strongly Agree |
|---|---|---|---|---|
| -2 | -1 | 0 | 1 | 2 |
| 55 | 15 | 14 | 5 | 2 |

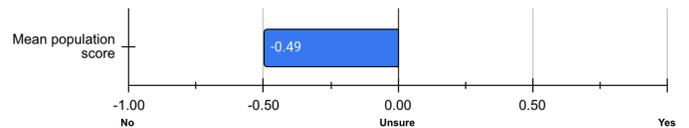

**Does having a conversation in an AI-powered chat feel natural to you?**

11. The mean population score of the Yes/No/Unsure question asking "Does having a conversation in an AI-powered chat feel natural to you?"

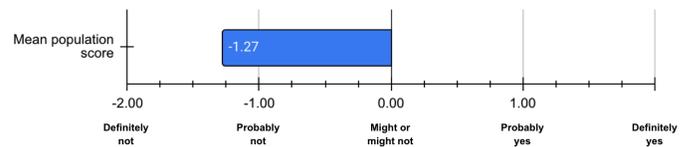

**Do you speak to a voice assistant in the same way you speak to a friend?**

12. The mean population score of the Likert scale question stating "Do you speak to a voice assistant in the same way you speak to a friend?"

## IV. DISCUSSION

The *Turing Test* has been popularized as a method of determining behavioral parity between Humans and AI. This test was first coined by Alan Turing and called "The Imitation Game" [11]. Essentially, if a human is unable to judge if interactions through a text interface are from a human or a machine, then the machine would pass the test. This test however, while may be a good indicator of general AI behavior, would be comparable to having a conversation with a stranger - someone for whom a theory of mind would not exist, or is in the early stages of developing. For those that experience social anxiety when interacting with complete strangers, they may experience a similar feeling when interacting with an AI agent as well.

In order to achieve a higher level of trust and predictable communication as we experience in human-to-human interactions, a theory of mind must be established [8, p. 252-260]. This is a level of interaction that the *Turing Test* would be unable to measure.

*A. The purpose of a theory of mind*

Through predictive processing, a theory of mind allows us to establish mental models that are constantly making predictions. Through case-based reasoning, these models are refined with both positive and negative cases over time [8, p. 258]. If we think about how we construct a theory of mind for people we interact with, and it gets refined over time the more we interact with someone, we have very robust theory of mind representations of our friends and people we know well [7].

These mental models contained within our theories of mind increase our comfort level with the types of interactions that they represent because we *can anticipate how people will respond*. This ability to anticipate responses indicates a strong theory of mind mental model, and is what makes communication with particular people feel natural and comfortable. This is why interacting with a close friend feels much more pleasant than interacting with someone who you cannot relate well to [8].

*B. Mutual theory of mind hindrance*

In traditional human-to-human communication, both humans in a given interaction are able to develop a theory of mind for each other. When you have two people who are friends and have highly refined theory of mind representations for each other, communication is fluid and effortless between both parties. This is called a mutual theory of mind. If we desire to interact with AI in a similar manner, not only do humans have to effectively develop a theory of mind representation for the AI agent, but AI agents also have to develop a theory of mind for humans.

Arguably, it could be easier for a theory of mind of a human to develop in AI than a theory of mind of an AI to develop in a human [4]. AI agents could also be extended to receive and interpret inputs from the multiple "interfaces" humans usually use to communicate (i.e. inflection, body language, facial expressions). An AI agent inherently does not have the ability to communicate using these methods, meaning that humans have much less information to build a theory of mind.

AI is commonly described as a black box. There are so many parameters that go into training an AI agent that nobody is really sure how an AI agent arrived at the response that it did. This nuance is one of the limiting factors that prevents us from being able to develop a theory of mind for AI agents, and is a vast inhibitor to establishing a mutual theory of mind [2], [6]. Explainable AI or Transparent AI may be an avenue to overcome this in future research [5], [6].

*C. Implications of survey findings*

These survey findings and themes have implications in the field of Human-AI Interaction which is an emerging sub-specialty of Human-Computer Interaction (HCI).

1) *Attempting to establish a theory of mind for an AI agent:* Participants intentionally alter their communication styles when interacting with AI. This alteration suggests that humans do instinctively attempt to construct a theory of mind representation even for Artificial Intelligence agents. In the case of voice assistants, people shorten sentences and enunciate more clearly. This *may* be because their theory of mind representation tells them that voice assistants have difficulty with complex sentences and parsing spoken language.

2) *Success or failure to establish a theory of mind for an AI agent:* People interact with AIs differently than with friends, but also in a way that does not seem natural. Keeping in mind that there is some evidence that a theory of mind may exist *to an extent*, the consensus that it still does not feel natural suggests that we have difficulty establishing a theory of mind representation for AIs [2], [9]. AI's limited communication interfaces (i.e. unable to use body language, et. cetera) combined with the opaque nature of Artificial Intelligence with current technology, the survey results provide no indication that robust theory of mind representations have been successfully established with the AI agents that respondents have interacted with.

3) *Cognitive load concerns:* The difficulty in establishing a theory of mind means that people will not be having natural conversations with AI agents. When these conversations are happening, human participants will be continuously attempting to refine their theory of mind while simultaneously engaging in conversation. This is comparable to the extra effort needed in conversations with unfamiliar people - but this never improves over time. This refinement of the theory of mind results in substantially more information being processed than communicating using an existing robust theory of mind representation. In turn, added information processing will also increase cognitive load [3], [10].

*D. Considerations for the future of Human-AI Interaction*

It may be advantageous to end users for new software products to limit the usage of natural language interfaces until AI technology progresses to better support theory of mind development (through the development of Explainable AI or other breakthroughs). AI could potentially still be used. Providing interactions with AI to power and inform traditional software interfaces may prevent users from attempting to develop a theory of mind representation for the AI agent, and thus avoid increasing cognitive load and limit the amount of information users must actively process.

*E. Considerations for technologists currently implementing AI*

The technology to allow for people to reliably establish theory of mind representations of AI still requires significant research and breakthroughs. Because of this, it is likely that we will continue to live for some time in an age of AI without the ability to fully develop a theory of mind, if not in perpetuity. Key takeaways from this study that can be readily applied to technology today include:

1) *Explore alternative interfaces:* To be clear, AI is not necessarily what leads to increased cognitive load - the natural language interface does. Exploring alternative interfaces may be an effective approach to still use AI in a software product without triggering a user's innate attempt to create a theory of mind mental model. Using AI to assist in things like data analysis and trend identification is a powerful way to use AI without a natural language interface component.

2) *When using natural language interfaces, emphasize predictability:* Natural language interfaces will likely remain an inevitable part of technology products for the foreseeable future. If and when natural language interfaces must be used, technologists should emphasize and make concerted efforts to make the AI agents behind them as predictable as possible. While a full theory of mind representation might still not be developed from creating a predictable AI agent, it may still allow users to learn to use it as a tool - like any other software product. This would theoretically allow the cognitive load required to use the natural language interface to decrease with usage.

## V. Conclusion

The survey data collected during the research process is clear, humans do indeed alter communication style *and* behavior when interacting with natural language interfaces. Through external literature review these behavioral changes and added difficulty can be attributed to difficulty establishing theory of mind representations for the AI agents that we interact with.

It seems that humans attempt to establish a theory of mind representation for an AI, but due to the nature of natural language interfaces it is, at best, extremely difficult to establish. At worst, humans may fail to establish a theory of mind representation for an AI altogether.

The result is an experience comparable to interacting with a complete stranger. This can result in an ineffective communication attempt and potentially even social anxiety.

There is ongoing research in the field of AI to determine the best way to innovate on abilities to establish methods for humans to develop theory of mind representations for AI agents; however, the technology is simply not where it needs to be yet in order to support this successfully.

## VI. Limitations & Future Work

This study was conducted asynchronously using a survey and various digital distribution methods. As such, all data recorded in this study is self-reported and not based on direct observation of participants placed into specific scenarios. This topic would benefit from a more robust empirical study in a controlled environment and exposing participants to specific interactions with AI and natural language interfaces. This would provide the opportunity to observe and discuss reactions with participants in real time.

*A. Future research opportunities*

It is worth exploring the potential for social anxiety when interacting with an AI for those that also experience it in typical circumstances. If this phenomenon does in fact exist, it may indicate that social anxiety could also be attributed to limited theory of mind development. This has the potential to have further implications in Human-AI Interaction best practices, accessibility and universal design principles, as well as emerging general/social anxiety research.

More research is also needed to determine the specific effects that interacting with an AI has on cognitive load. While the hypothesis that working with an immature theory of mind or without one entirely would increase cognitive load is sound theoretically, the expected increase compared to completing the same task using traditional software interfaces is unknown and has yet to be measured.

APPENDIX

*1. Survey questions and results*

Free response and segmentation questions are withheld from the following report to protect the privacy of participants. Redacted versions of these sections can be made available upon request.

The rest of this page is intentionally left blank. The survey questions begin on the next page.

# Communication with technology / Familiarity

Responses: 101

Have you ever interacted with any form of artificial intelligence? (ChatGPT, voice assistant, automated customer service call center)  95

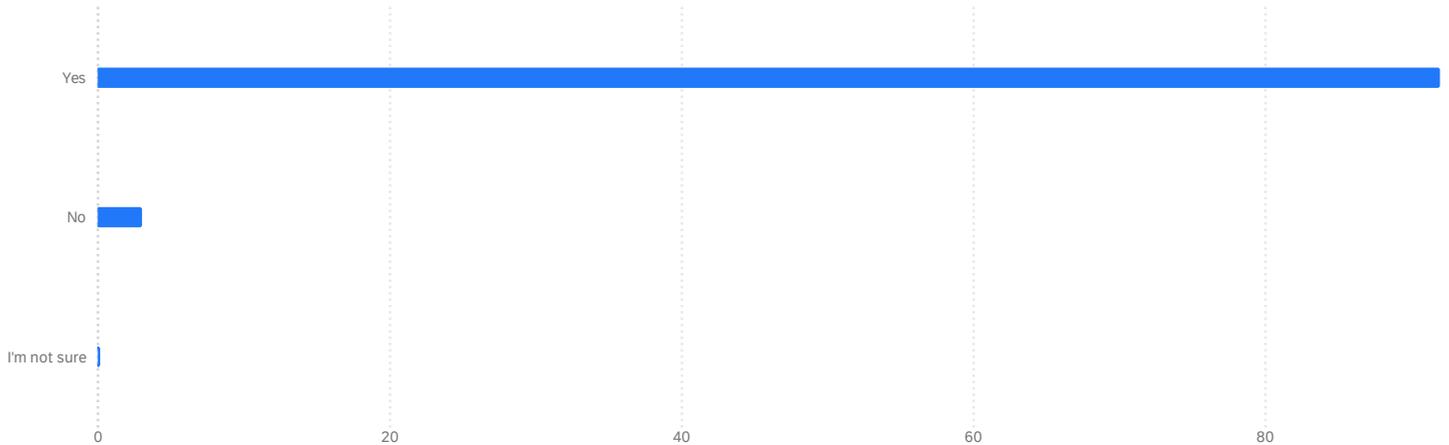

Have you ever interacted with any form of artificial intelligence? (ChatGPT, voice assistant, automated customer service call center)  95

| Q2.1 - Have you ever interacted with any form of artificial intelligence? (ChatGPT, voice assistant, automated customer service call center) | Percentage | Count |
| --- | --- | --- |
| Yes | 97% | 92 |
| No | 3% | 3 |
| I'm not sure | 0% | 0 |
| Sum | 100% | 95 |

Have you ever used an AI-powered chat such as ChatGPT, or a customer support chat that started with an AI-powered digital assistant?  95

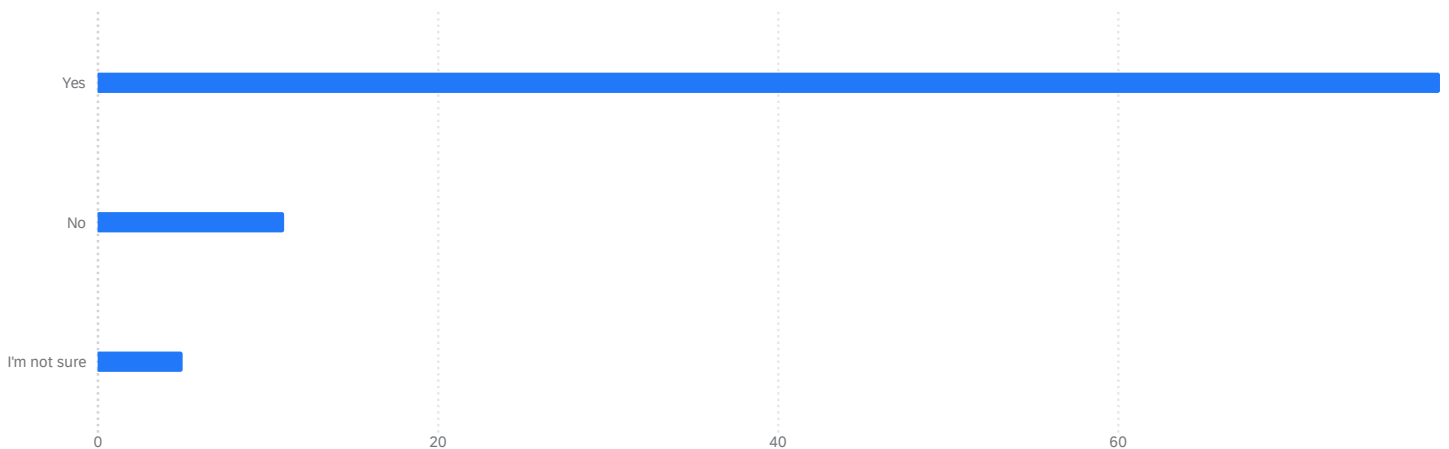

## Have you ever used an AI-powered chat such as ChatGPT, or a customer support chat that started with an AI-powered digital assistant? 95

| Q2.2 - Have you ever used an AI-powered chat such as ChatGPT, or a customer support chat that started with an AI-powered digital assistant? | Percentage | Count |
|---|---|---|
| Yes | 83% | 79 |
| No | 12% | 11 |
| I'm not sure | 5% | 5 |
| Sum | 100% | 95 |

## How comfortable are you with using technology in general? 94

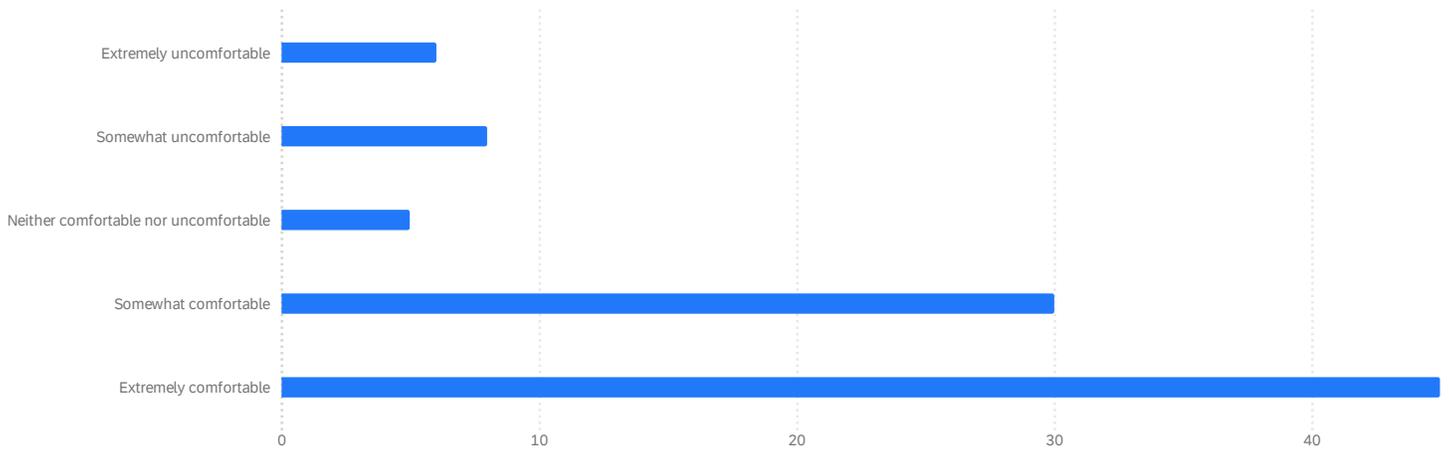

## How comfortable are you with using technology in general? 94

| Q3.1 - How comfortable are you with using technology in general? | Percentage | Count |
|---|---|---|
| Extremely uncomfortable | 6% | 6 |
| Somewhat uncomfortable | 9% | 8 |
| Neither comfortable nor uncomfortable | 5% | 5 |
| Somewhat comfortable | 32% | 30 |
| Extremely comfortable | 48% | 45 |
| Sum | 100% | 94 |

How likely are you to begin using new technology as soon as it becomes available? 94

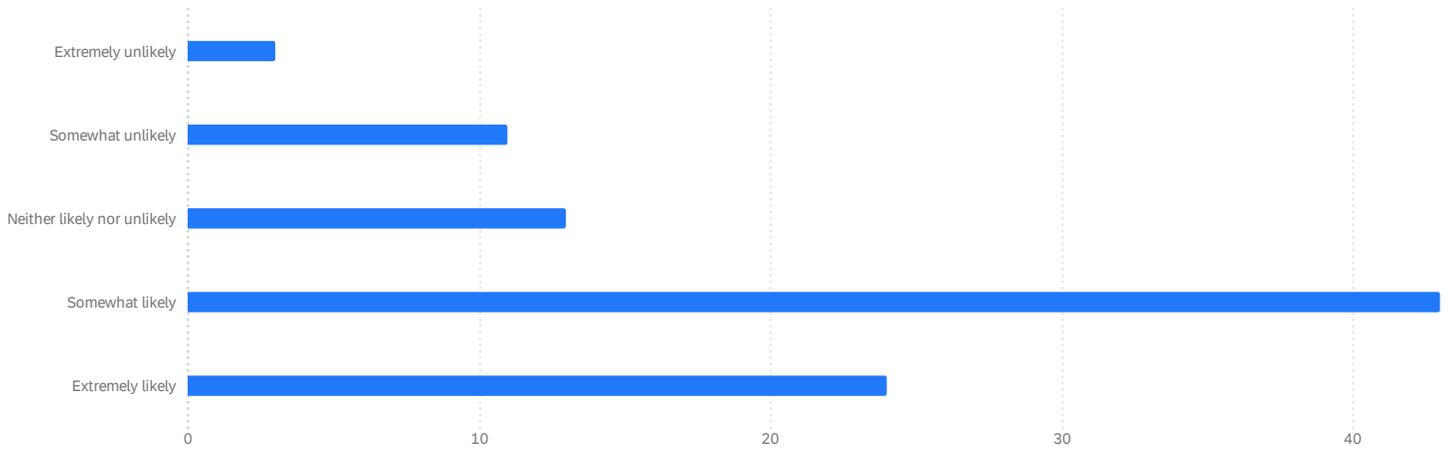

How likely are you to begin using new technology as soon as it becomes available? 94

| Q3.2 - How likely are you to begin using new technology as soon as it becomes available? | Percentage | Count |
| --- | --- | --- |
| Extremely unlikely | 3% | 3 |
| Somewhat unlikely | 12% | 11 |
| Neither likely nor unlikely | 14% | 13 |
| Somewhat likely | 46% | 43 |
| Extremely likely | 26% | 24 |
| Sum | 100% | 94 |

How familiar are you with Artificial Intelligence as a concept? 94

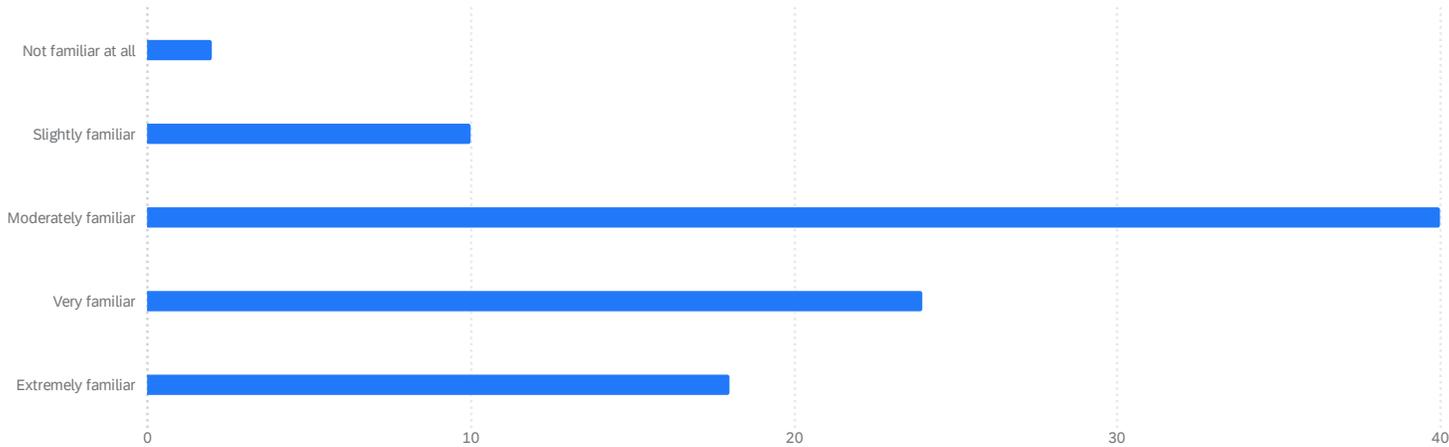

## How familiar are you with Artificial Intelligence as a concept? 94

| Q3.3 - How familiar are you with Artificial Intelligence as a concept? | Percentage | Count |
| --- | --- | --- |
| Not familiar at all | 2% | 2 |
| Slightly familiar | 11% | 10 |
| Moderately familiar | 43% | 40 |
| Very familiar | 26% | 24 |
| Extremely familiar | 19% | 18 |
| Sum | 100% | 94 |

## How familiar are you with using Artificial Intelligence-powered products and software? 94

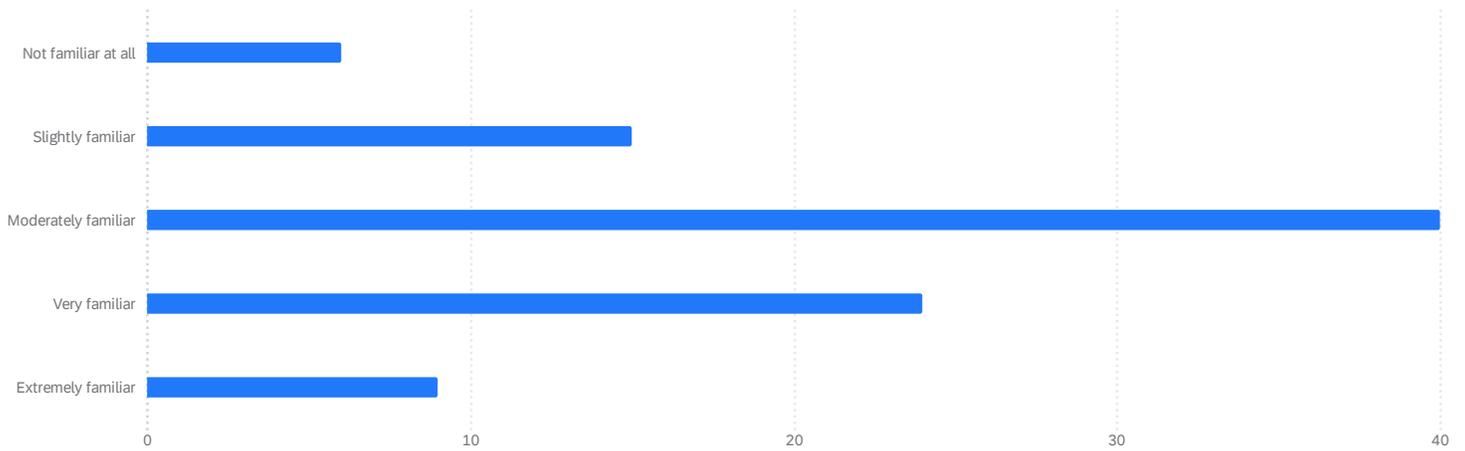

## How familiar are you with using Artificial Intelligence-powered products and software? 94

| Q3.4 - How familiar are you with using Artificial Intelligence-powered products and software? | Percentage | Count |
| --- | --- | --- |
| Not familiar at all | 6% | 6 |
| Slightly familiar | 16% | 15 |
| Moderately familiar | 43% | 40 |
| Very familiar | 26% | 24 |
| Extremely familiar | 10% | 9 |
| Sum | 100% | 94 |

In general, how much do you trust Artificial Intelligence? 94

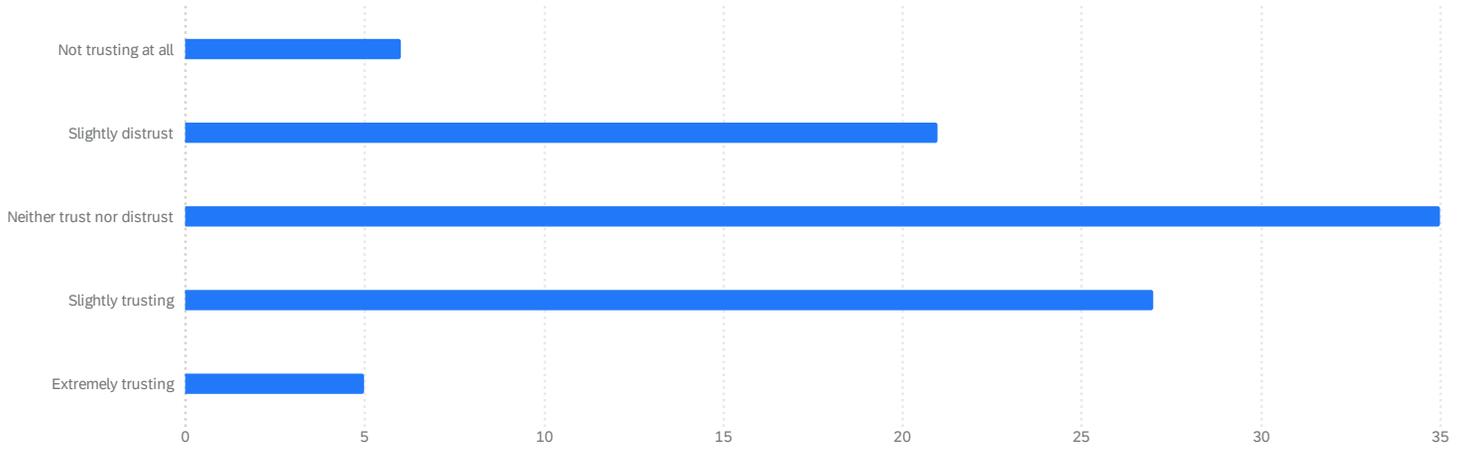

In general, how much do you trust Artificial Intelligence? 94

| Q3.5 - In general, how much do you trust Artificial Intelligence? | Percentage | Count |
| --- | --- | --- |
| Not trusting at all | 6% | 6 |
| Slightly distrust | 22% | 21 |
| Neither trust nor distrust | 37% | 35 |
| Slightly trusting | 29% | 27 |
| Extremely trusting | 5% | 5 |
| Sum | 100% | 94 |

# Communication with technology / Frequency

Responses: 101

How often do you use voice assistants? 93

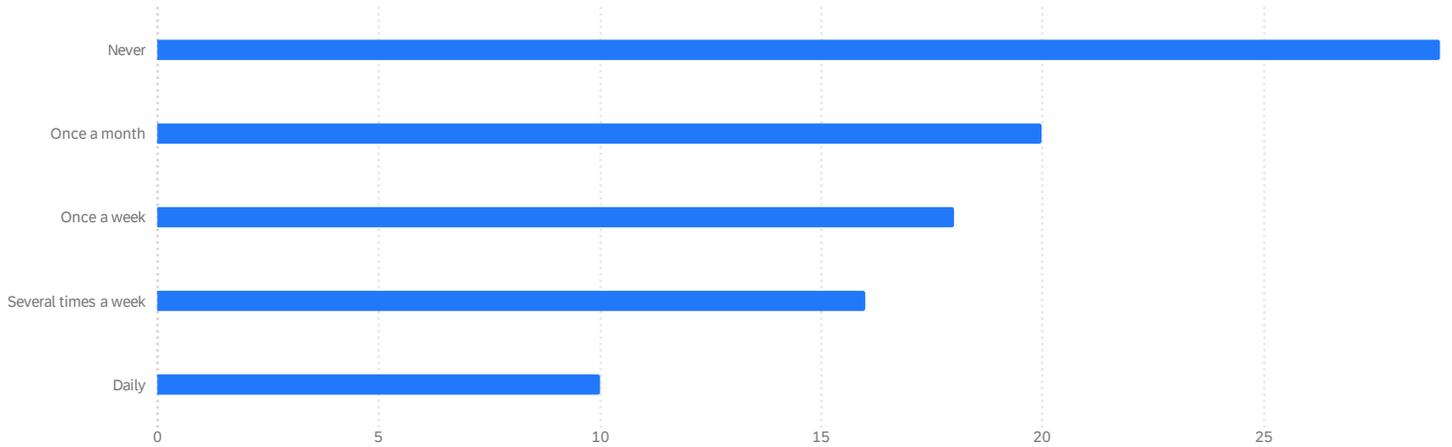

How often do you use voice assistants? 93

| Q4.1 - How often do you use voice assistants? | Percentage | Count |
| --- | --- | --- |
| Never | 31% | 29 |
| Once a month | 22% | 20 |
| Once a week | 19% | 18 |
| Several times a week | 17% | 16 |
| Daily | 11% | 10 |
| Sum | 100% | 93 |

How often do you use an AI-powered chat? (This could be customer support, or a program like ChatGPT) 93

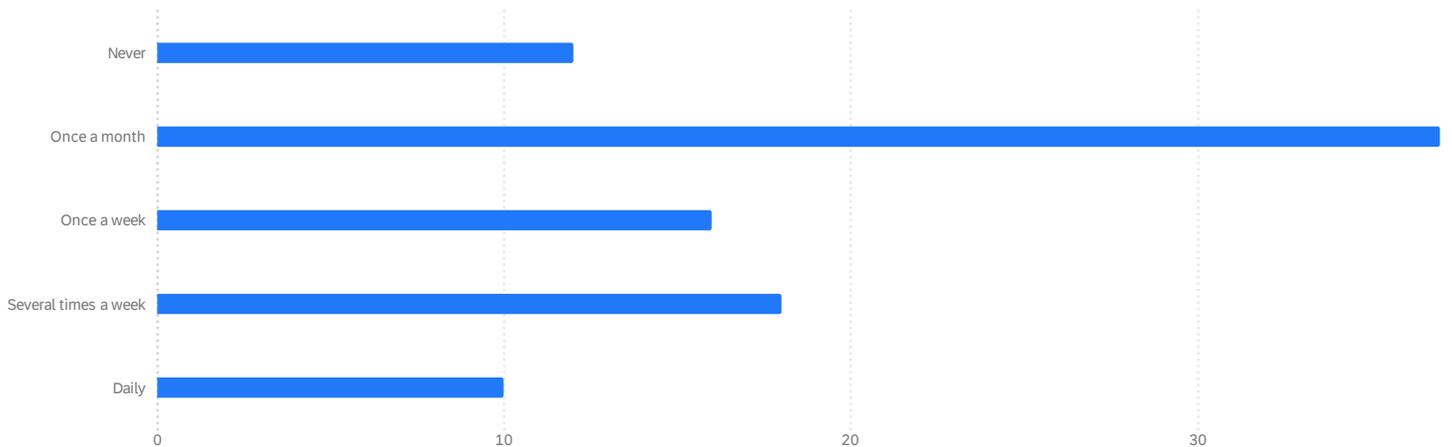

How often do you use an AI-powered chat? (This could be customer support, or a program like ChatGPT) 93

| Q4.2 - How often do you use an AI-powered chat? (This could be customer support, or a program like ChatGPT) | Percentage | Count |
|---|---|---|
| Never | 13% | 12 |
| Once a month | 40% | 37 |
| Once a week | 17% | 16 |
| Several times a week | 19% | 18 |
| Daily | 11% | 10 |
| Sum | 100% | 93 |

How frequently do you plan out individual words in a sentence when speaking to or messaging with a friend? 91

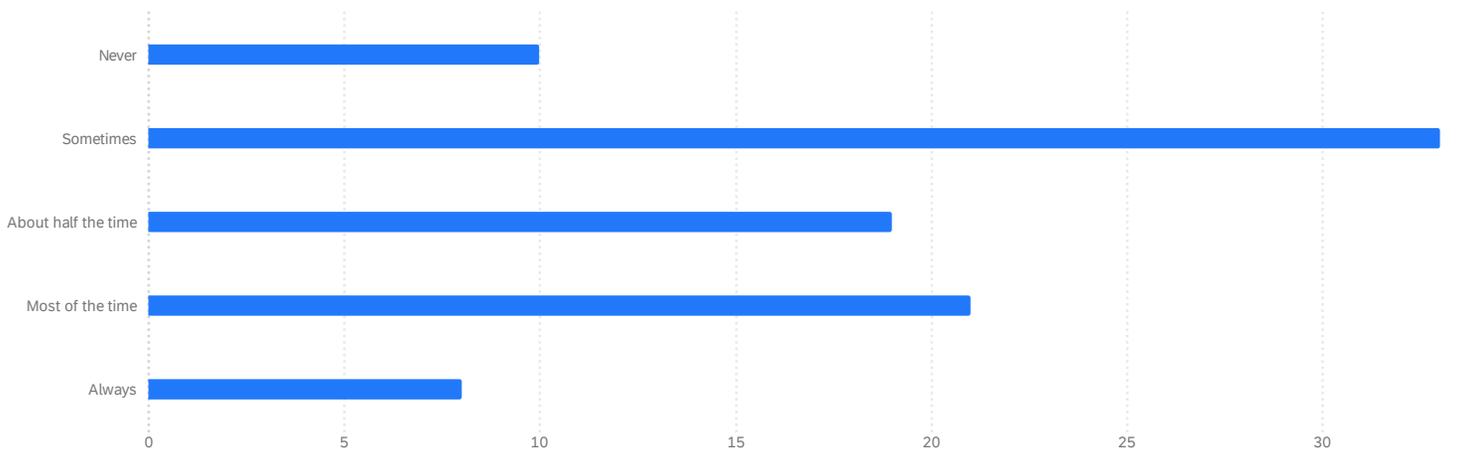

How frequently do you plan out individual words in a sentence when speaking to or messaging with a friend? 91

| Q5.2 - How frequently do you plan out individual words in a sentence when speaking to or messaging with a friend? | Percentage | Count |
|---|---|---|
| Never | 11% | 10 |
| Sometimes | 36% | 33 |
| About half the time | 21% | 19 |
| Most of the time | 23% | 21 |
| Always | 9% | 8 |
| Sum | 100% | 91 |

## How frequently do you need to repeat yourself when speaking to or messaging with a friend? 91

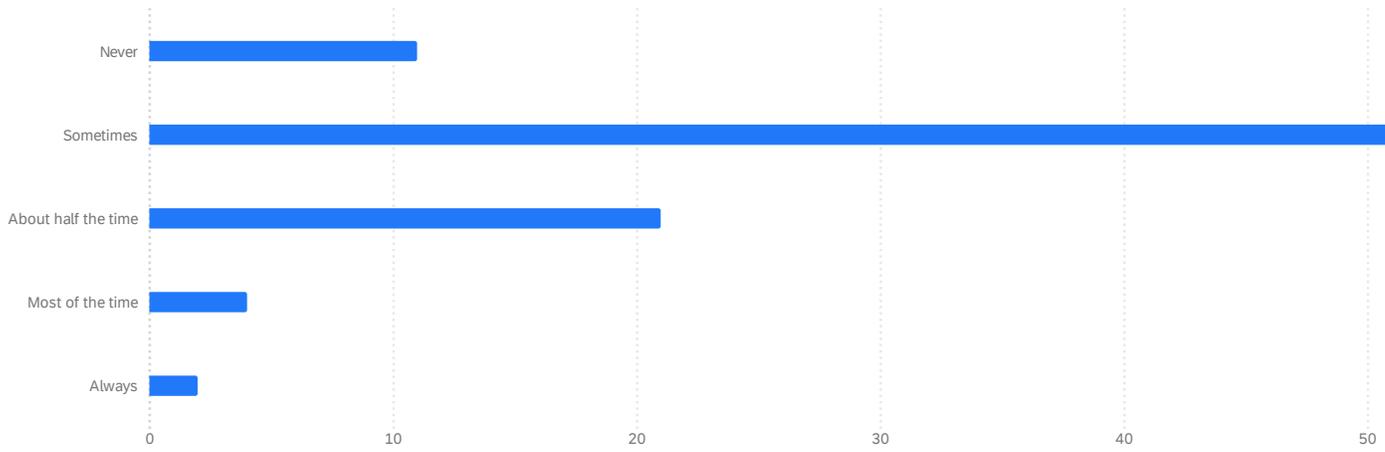

## How frequently do you need to repeat yourself when speaking to or messaging with a friend? 91

| Q5.3 - How frequently do you need to repeat yourself when speaking to or messaging with a friend? | Percentage | Count |
| --- | --- | --- |
| Never | 12% | 11 |
| Sometimes | 58% | 53 |
| About half the time | 23% | 21 |
| Most of the time | 4% | 4 |
| Always | 2% | 2 |
| Sum | 100% | 91 |

## How frequently do you need to rephrase something you said when speaking to or messaging with a friend? 91

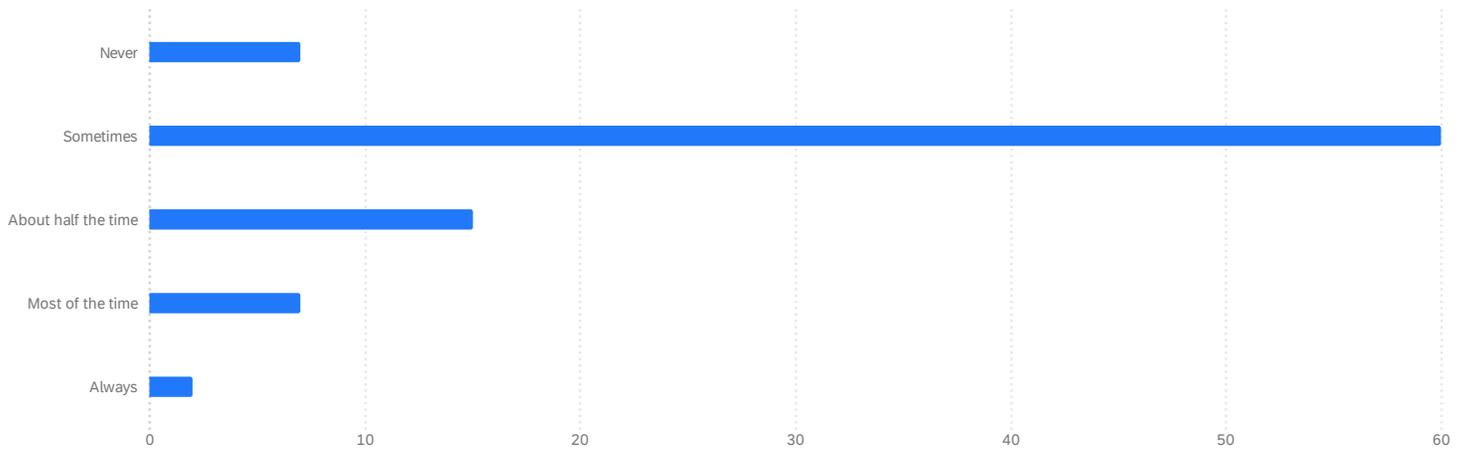

How frequently do you need to rephrase something you said when speaking to or messaging with a friend? 91 ⓘ

| Q5.4 - How frequently do you need to rephrase something you said when speaking to or messaging with a friend? | Percentage | Count |
| --- | --- | --- |
| Never | 8% | 7 |
| Sometimes | 66% | 60 |
| About half the time | 16% | 15 |
| Most of the time | 8% | 7 |
| Always | 2% | 2 |
| Sum | 100% | 91 |

How frequently do you plan out individual words in a sentence when speaking to a voice assistant? 91 ⓘ

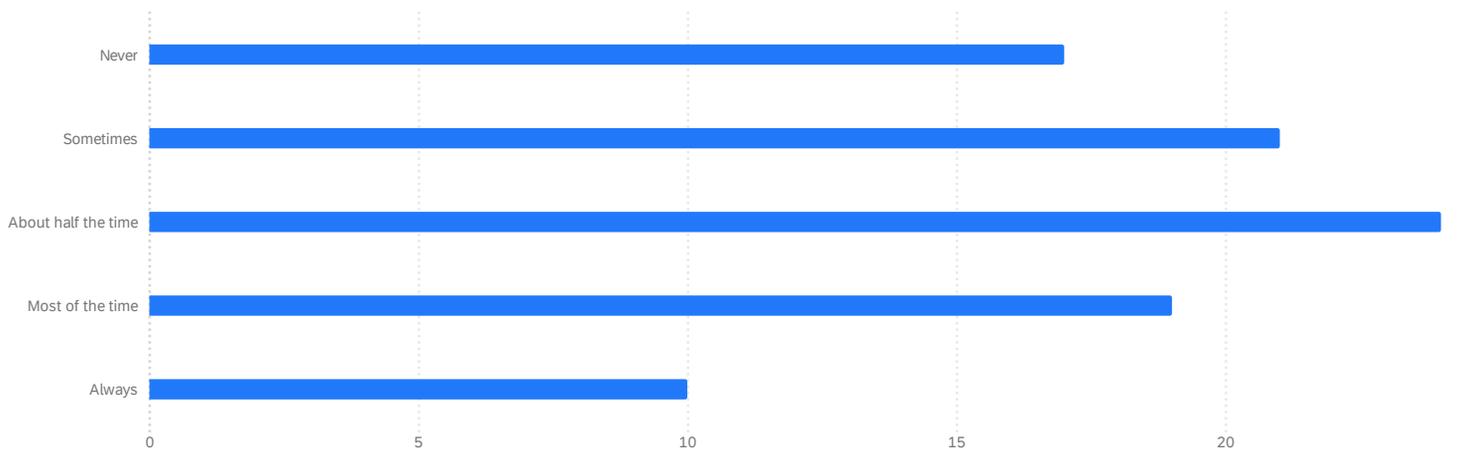

How frequently do you plan out individual words in a sentence when speaking to a voice assistant? 91 ⓘ

| Q5.6 - How frequently do you plan out individual words in a sentence when speaking to a voice assistant? | Percentage | Count |
| --- | --- | --- |
| Never | 19% | 17 |
| Sometimes | 23% | 21 |
| About half the time | 26% | 24 |
| Most of the time | 21% | 19 |
| Always | 11% | 10 |
| Sum | 100% | 91 |

How frequently do you need to repeat yourself when speaking to a voice assistant? 91

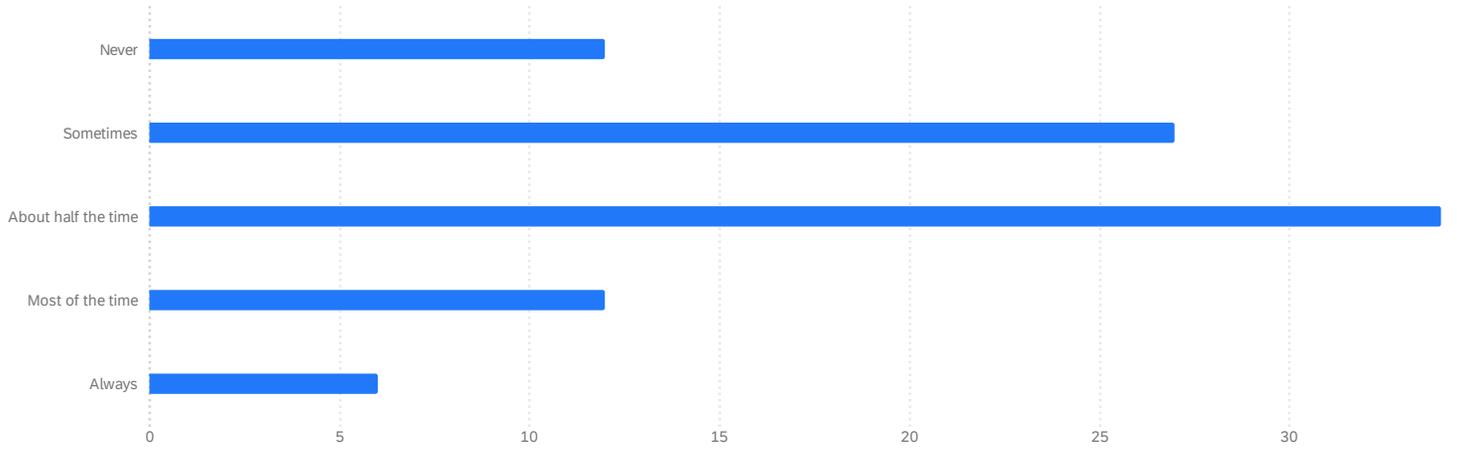

How frequently do you need to repeat yourself when speaking to a voice assistant? 91

| Q5.7 - How frequently do you need to repeat yourself when speaking to a voice assistant? | Percentage | Count |
| --- | --- | --- |
| Never | 13% | 12 |
| Sometimes | 30% | 27 |
| About half the time | 37% | 34 |
| Most of the time | 13% | 12 |
| Always | 7% | 6 |
| Sum | 100% | 91 |

How frequently do you need to rephrase something you said when speaking to a voice assistant? 91

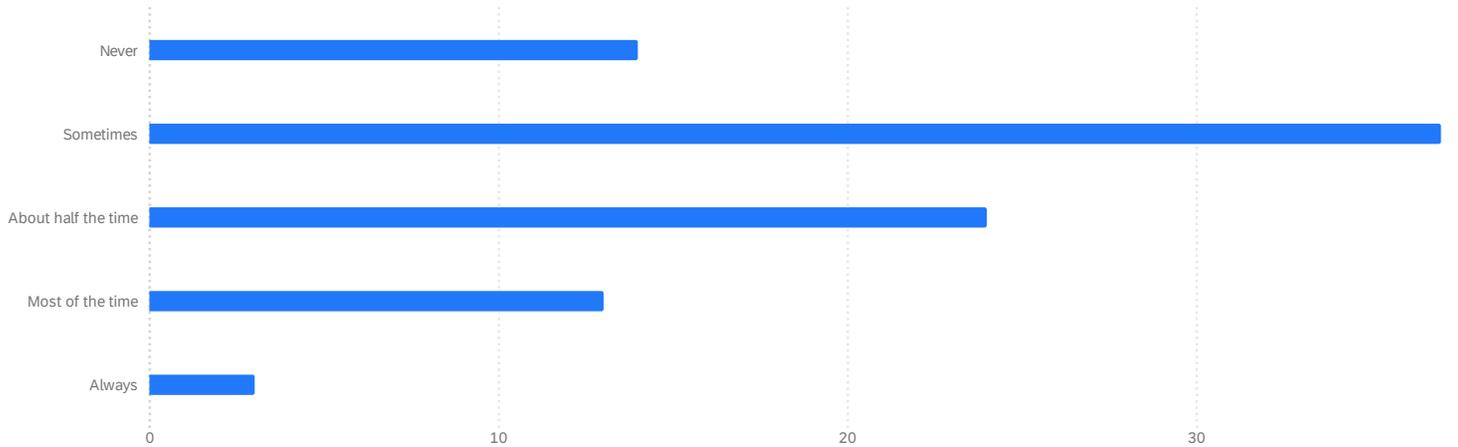

How frequently do you need to rephrase something you said when speaking to a voice assistant? 91

| Q5.8 - How frequently do you need to rephrase something you said when speaking to a voice assistant? | Percentage | Count |
| --- | --- | --- |
| Never | 15% | 14 |
| Sometimes | 41% | 37 |
| About half the time | 26% | 24 |
| Most of the time | 14% | 13 |
| Always | 3% | 3 |
| Sum | 100% | 91 |

How frequently do you plan out individual words in a sentence when using an AI-powered chat? 91

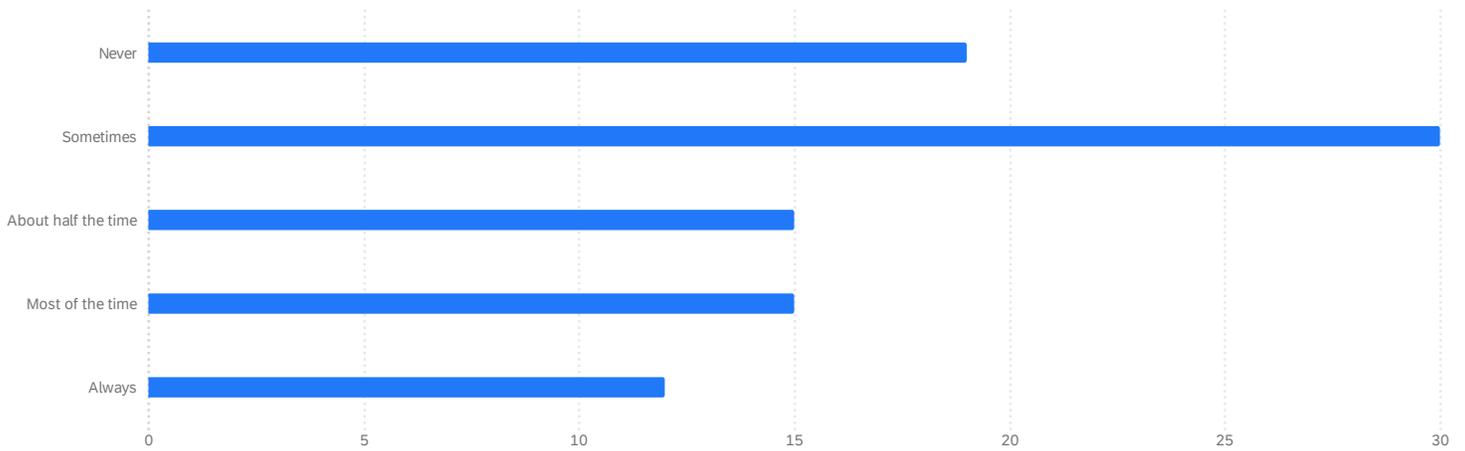

How frequently do you plan out individual words in a sentence when using an AI-powered chat? 91

| Q5.10 - How frequently do you plan out individual words in a sentence when using an AI-powered chat? | Percentage | Count |
| --- | --- | --- |
| Never | 21% | 19 |
| Sometimes | 33% | 30 |
| About half the time | 16% | 15 |
| Most of the time | 16% | 15 |
| Always | 13% | 12 |
| Sum | 100% | 91 |

How frequently do you need to repeat yourself when using an AI-powered chat? 91

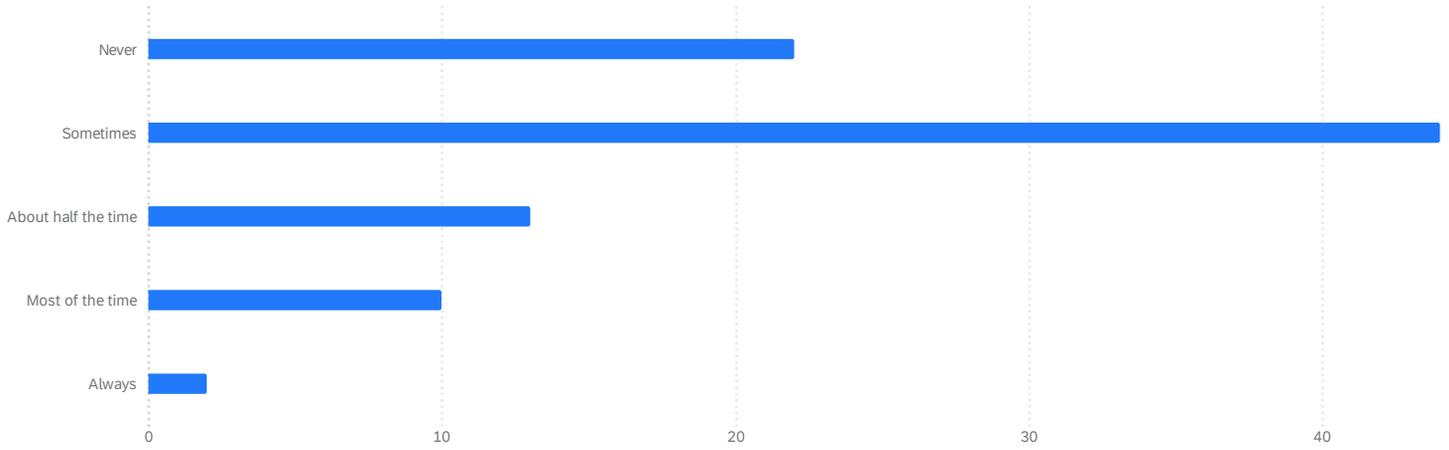

How frequently do you need to repeat yourself when using an AI-powered chat? 91

| Q5.11 - How frequently do you need to repeat yourself when using an AI-powered chat? | Percentage | Count |
| --- | --- | --- |
| Never | 24% | 22 |
| Sometimes | 48% | 44 |
| About half the time | 14% | 13 |
| Most of the time | 11% | 10 |
| Always | 2% | 2 |
| Sum | 100% | 91 |

How frequently do you need to rephrase something you said when using an AI-powered chat? 91

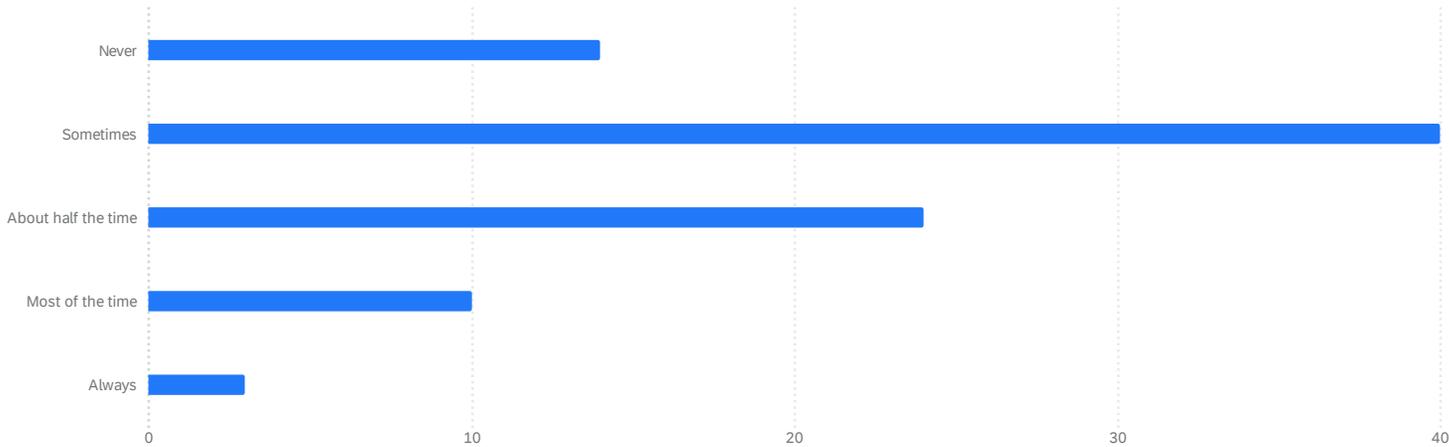

### How frequently do you need to rephrase something you said when using an AI-powered chat? 91

| Q5.12 - How frequently do you need to rephrase something you said when using an AI-powered chat? | Percentage | Count |
|---|---|---|
| Never | 15% | 14 |
| Sometimes | 44% | 40 |
| About half the time | 26% | 24 |
| Most of the time | 11% | 10 |
| Always | 3% | 3 |
| Sum | 100% | 91 |

### How frequently do you need to rephrase something you said when using an AI-powered chat? 91

| How frequently do you need to rephrase something you said when using an AI-... | Average | Minimum | Maximum | Count |
|---|---|---|---|---|
| Never | 1.00 | 1.00 | 1.00 | 14 |
| Sometimes | 2.00 | 2.00 | 2.00 | 40 |
| About half the time | 3.00 | 3.00 | 3.00 | 24 |
| Most of the time | 4.00 | 4.00 | 4.00 | 10 |
| Always | 5.00 | 5.00 | 5.00 | 3 |

# Communication with technology / Confidence

Responses: 101

When using a voice assistant (such as Siri, Google Assistant, Alexa), how confident are you that you will be heard accurately? 91

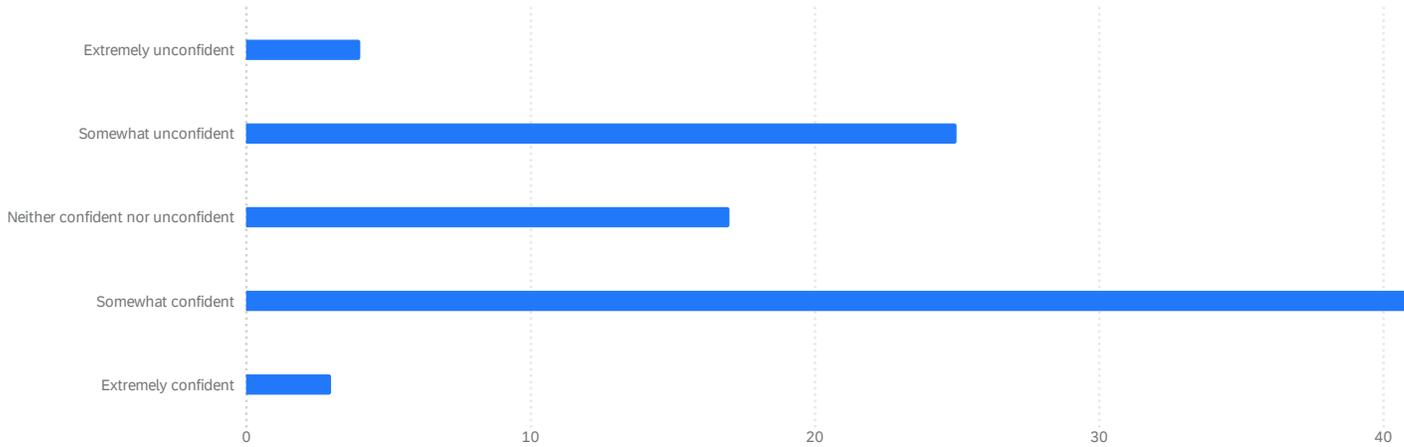

When using a voice assistant (such as Siri, Google Assistant, Alexa), how confident are you that you will be heard accurately? 91

| Q6.1 - When using a voice assistant (such as Siri, Google Assistant, Alexa), how confident are you that you will be heard accurately? | Percentage | Count |
| --- | --- | --- |
| Extremely unconfident | 4% | 4 |
| Somewhat unconfident | 27% | 25 |
| Neither confident nor unconfident | 19% | 17 |
| Somewhat confident | 46% | 42 |
| Extremely confident | 3% | 3 |
| Sum | 100% | 91 |

When using a voice assistant (such as Siri, Google Assistant, Alexa), how confident are you that you will be understood accurately? 91

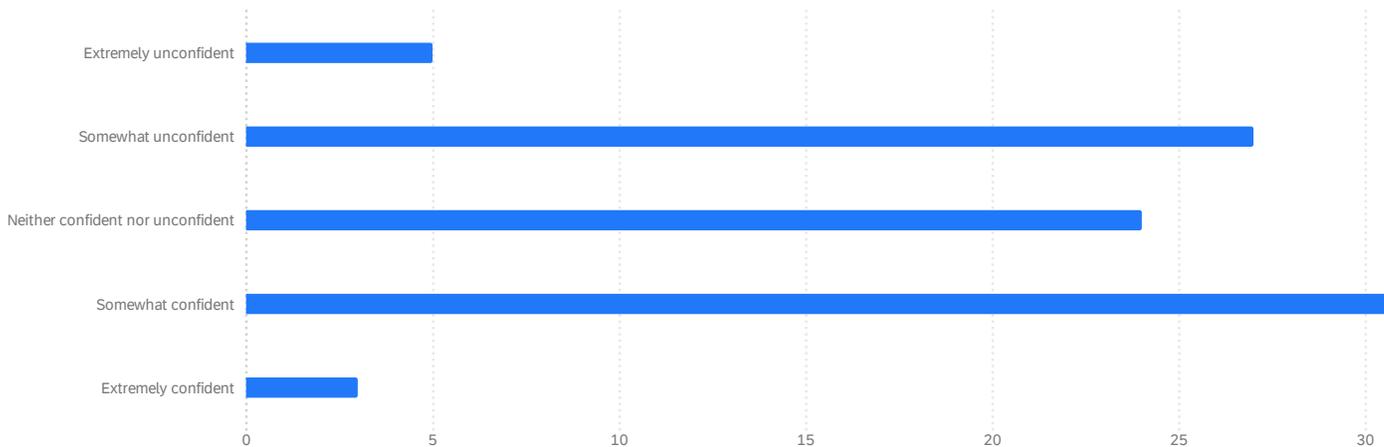

When using a voice assistant (such as Siri, Google Assistant, Alexa), how confident are you that you will be understood accurately? 91

| Q6.2 - When using a voice assistant (such as Siri, Google Assistant, Alexa), how confident are you that you will be understood accurately? | Percentage | Count |
|---|---|---|
| Extremely unconfident | 5% | 5 |
| Somewhat unconfident | 30% | 27 |
| Neither confident nor unconfident | 26% | 24 |
| Somewhat confident | 35% | 32 |
| Extremely confident | 3% | 3 |
| Sum | 100% | 91 |

When calling a customer support center that has an automated voice assistant start the call, how confident are you that it will correctly hear the words you say? 91

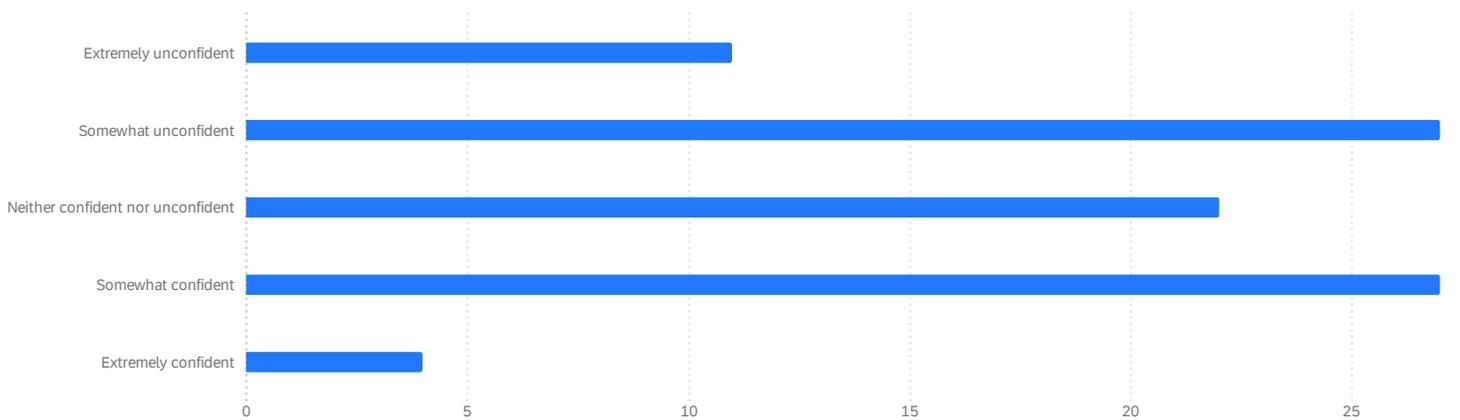

When calling a customer support center that has an automated voice assistant start the call, how confident are you that it will correctly hear the words you say? 91

| Q6.3 - When calling a customer support center that has an automated voice assistant start the call, how confident are you that it will correctly hear the words you say? | Percentage | Count |
|---|---|---|
| Extremely unconfident | 12% | 11 |
| Somewhat unconfident | 30% | 27 |
| Neither confident nor unconfident | 24% | 22 |
| Somewhat confident | 30% | 27 |
| Extremely confident | 4% | 4 |
| Sum | 100% | 91 |

When calling a customer support center that has an automated voice assistant start the call, how confident are you that it will correctly understand your comment or request? 91

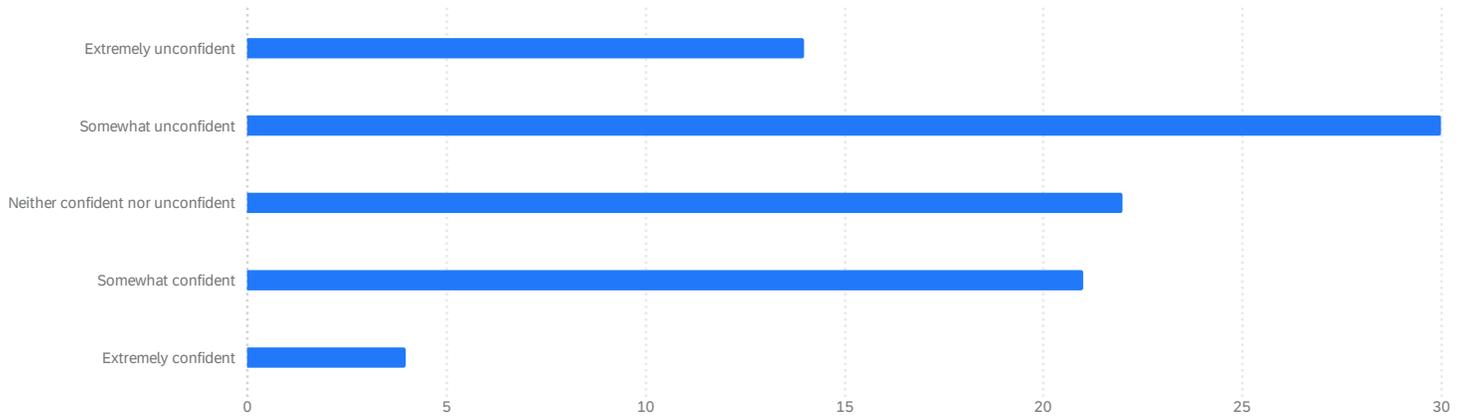

When calling a customer support center that has an automated voice assistant start the call, how confident are you that it will correctly understand your comment or request? 91

| Q6.4 - When calling a customer support center that has an automated voice assistant start the call, how confident are you that it will correctly understand your comment or request? | Percentage | Count |
| --- | --- | --- |
| Extremely unconfident | 15% | 14 |
| Somewhat unconfident | 33% | 30 |
| Neither confident nor unconfident | 24% | 22 |
| Somewhat confident | 23% | 21 |
| Extremely confident | 4% | 4 |
| Sum | 100% | 91 |

# Communication with technology / Behavior & Feelings

Responses: 101

Do you feel like you change the way you construct sentences when interacting with a voice assistant? 91

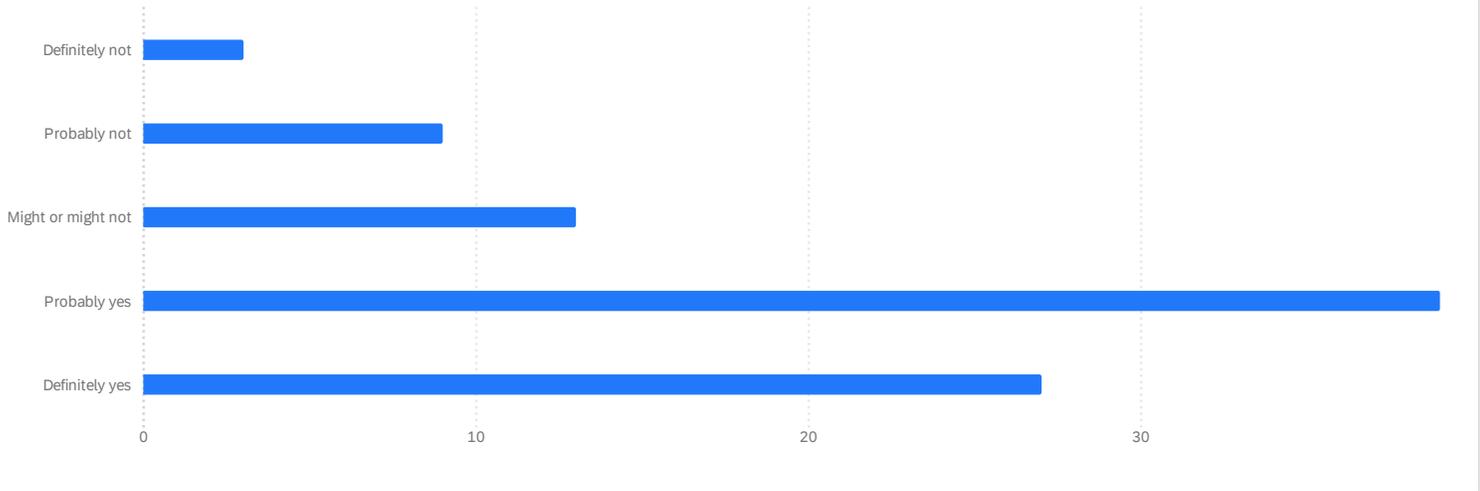

Do you feel like you change the way you construct sentences when interacting with a voice assistant? 91

| Q7.1 - Do you feel like you change the way you construct sentences when interacting with a voice assistant? | Percentage | Count |
| --- | --- | --- |
| Definitely not | 3% | 3 |
| Probably not | 10% | 9 |
| Might or might not | 14% | 13 |
| Probably yes | 43% | 39 |
| Definitely yes | 30% | 27 |
| Sum | 100% | 91 |

Do you feel like you change the way you construct sentences when interacting with an AI-powered chat? 91

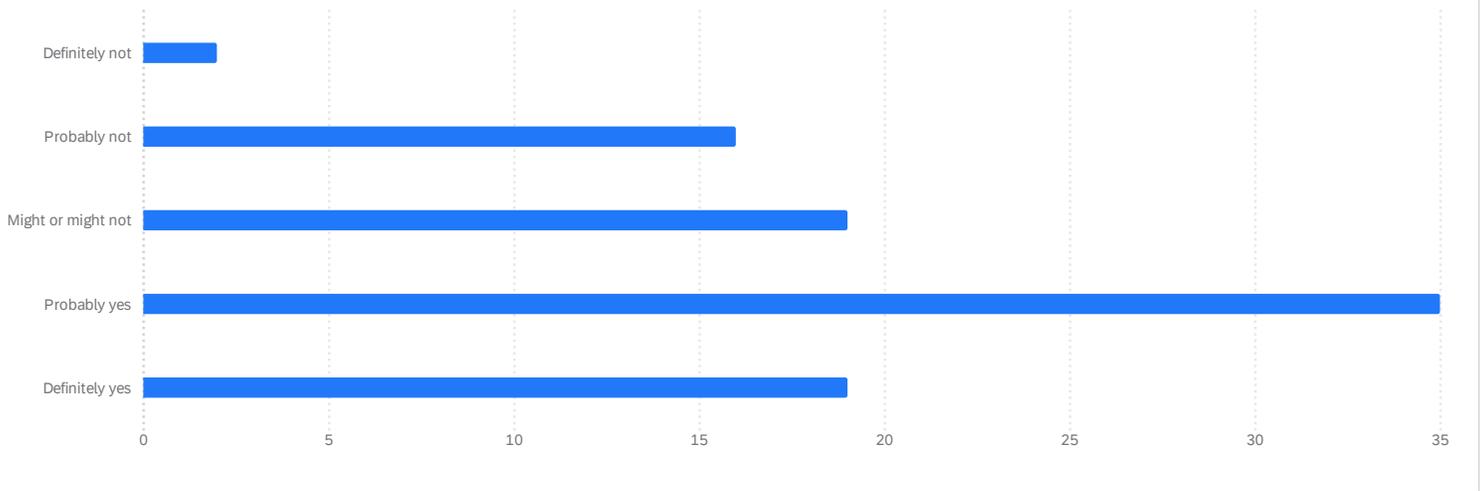

Do you feel like you change the way you construct sentences when interacting with an AI-powered chat? 91

| Q7.2 - Do you feel like you change the way you construct sentences when interacting with an AI-powered chat? | Percentage | Count |
|---|---|---|
| Definitely not | 2% | 2 |
| Probably not | 18% | 16 |
| Might or might not | 21% | 19 |
| Probably yes | 38% | 35 |
| Definitely yes | 21% | 19 |
| Sum | 100% | 91 |

Do you speak to a voice assistant in the same way you speak to a friend? 91

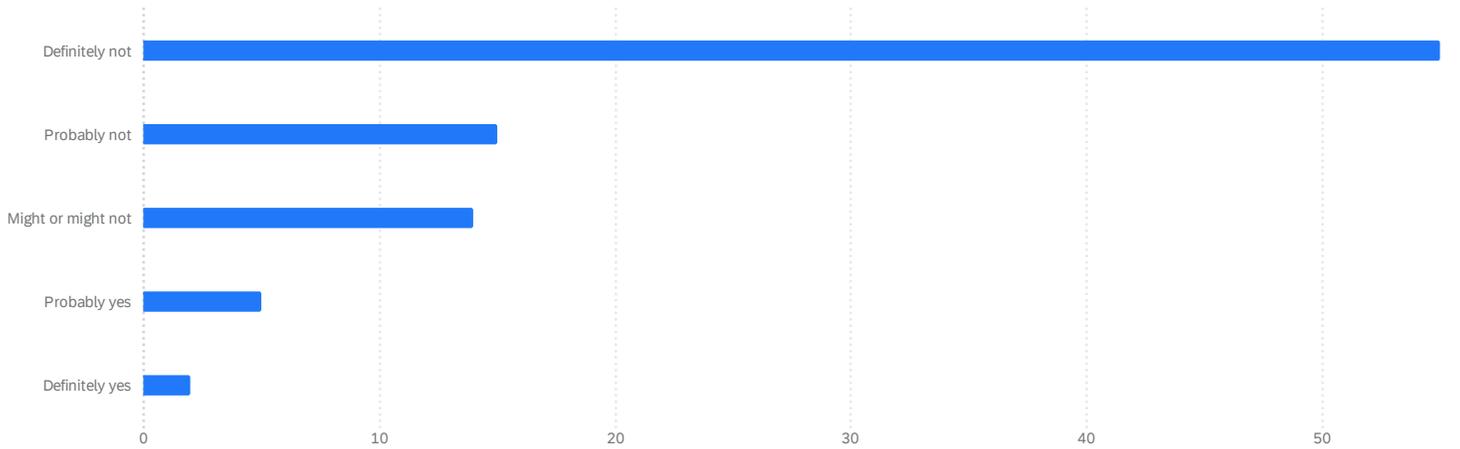

Do you speak to a voice assistant in the same way you speak to a friend? 91

| Q7.3 - Do you speak to a voice assistant in the same way you speak to a friend? | Percentage | Count |
|---|---|---|
| Definitely not | 60% | 55 |
| Probably not | 16% | 15 |
| Might or might not | 15% | 14 |
| Probably yes | 5% | 5 |
| Definitely yes | 2% | 2 |
| Sum | 100% | 91 |

Does speaking to a voice assistant feel natural to you? 91

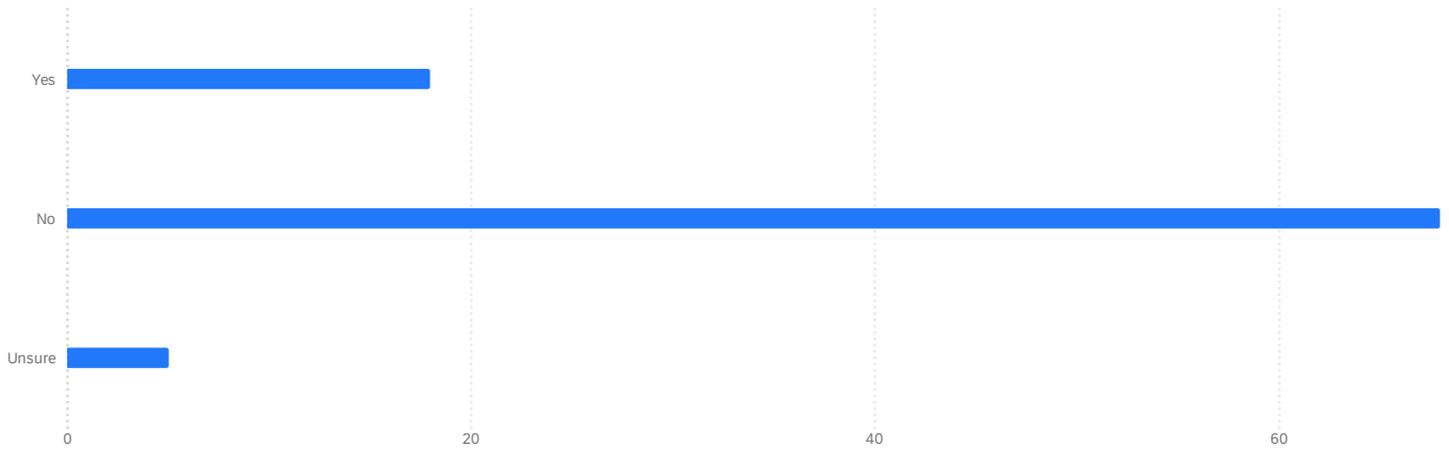

Does speaking to a voice assistant feel natural to you? 91

| Q7.4 - Does speaking to a voice assistant feel natural to you? | Percentage | Count |
| --- | ---: | ---: |
| Yes | 20% | 18 |
| No | 75% | 68 |
| Unsure | 5% | 5 |
| Sum | 100% | 91 |

Do you phrase sentences to an AI-powered chat conversation in the same way you phrase sentences in a message to a friend? 91

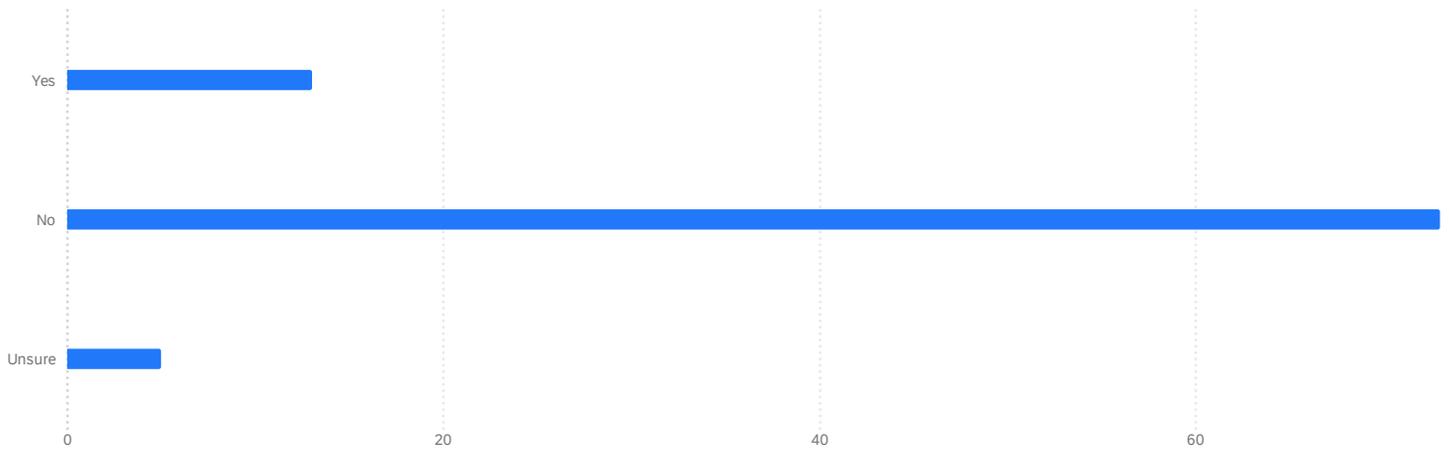

### Do you phrase sentences to an AI-powered chat conversation in the same way you phrase sentences in a message to a friend? 91

| Q7.5 - Do you phrase sentences to an AI-powered chat conversation in the same way you phrase sentences in a message to a friend? | Percentage | Count |
|---|---|---|
| Yes | 14% | 13 |
| No | 80% | 73 |
| Unsure | 5% | 5 |
| Sum | 100% | 91 |

### Does having a conversation in an AI-powered chat feel natural to you? 91

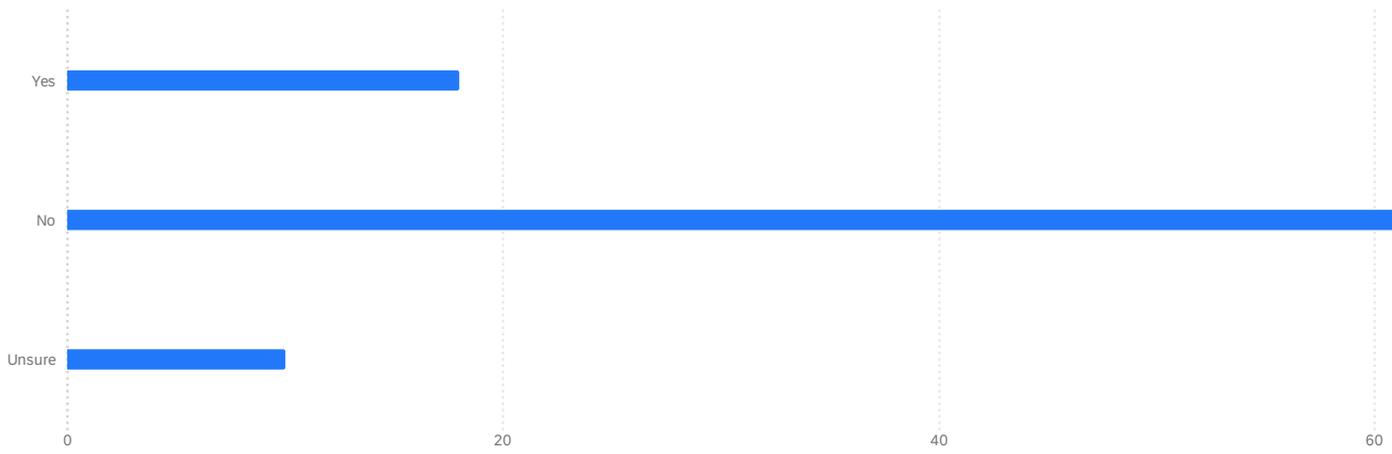

### Does having a conversation in an AI-powered chat feel natural to you? 91

| Q7.6 - Does having a conversation in an AI-powered chat feel natural to you? | Percentage | Count |
|---|---|---|
| Yes | 20% | 18 |
| No | 69% | 63 |
| Unsure | 11% | 10 |
| Sum | 100% | 91 |

### Does having a conversation in an AI-powered chat feel natural to you? 91

| Does having a conversation in an AI-powered chat feel natural to you? | Average | Minimum | Maximum | Count |
|---|---|---|---|---|
| Yes | 1.00 | 1.00 | 1.00 | 18 |
| No | 2.00 | 2.00 | 2.00 | 63 |
| Unsure | 3.00 | 3.00 | 3.00 | 10 |

### Does having a conversation with a voice assistant or AI-powered chat feel comparable to having a conversation with another person? 91

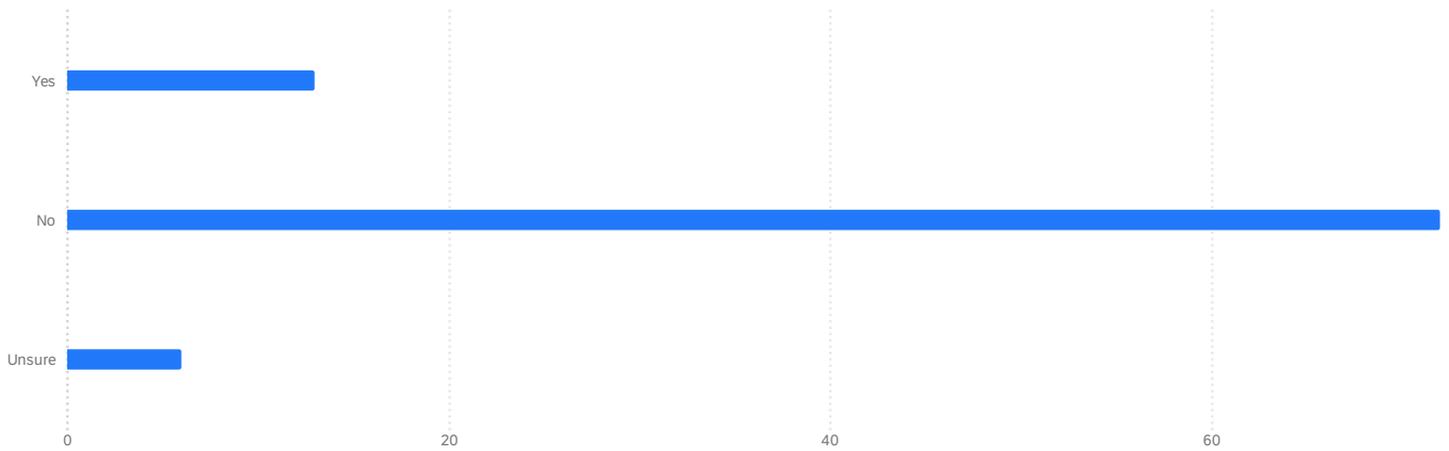

### Does having a conversation with a voice assistant or AI-powered chat feel comparable to having a conversation with another person? 91

| Q7.7 - Does having a conversation with a voice assistant or AI-powered chat feel comparable to having a conversation with another person? | Percentage | Count |
|---|---|---|
| Yes | 14% | 13 |
| No | 79% | 72 |
| Unsure | 7% | 6 |
| Sum | 100% | 91 |

Does having a conversation with a voice assistant or AI-powered chat feel comparable to having a conversation with another person? 91

| Does having a conversation with a voice assistant or AI-powered chat feel c... | Average | Minimum | Maximum | Count |
|---|---|---|---|---|
| Yes | 1.00 | 1.00 | 1.00 | 13 |
| No | 2.00 | 2.00 | 2.00 | 72 |
| Unsure | 3.00 | 3.00 | 3.00 | 6 |

# Communication with technology / Adaptation

Responses: 101

If you use voice interfaces, what aspects of your speech patterns change when using a voice assistant? 88

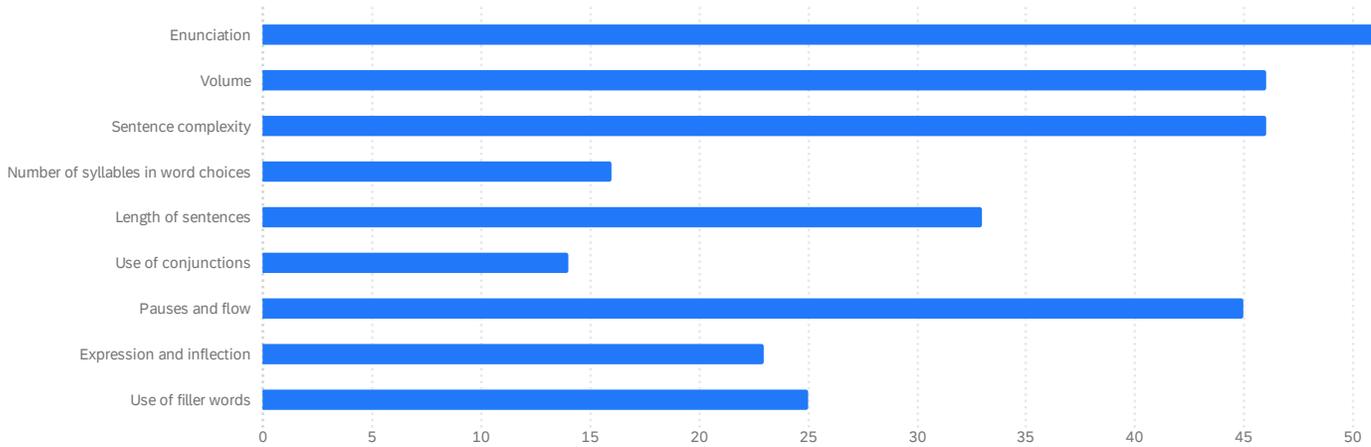

If you use voice interfaces, what aspects of your speech patterns change when using a voice assistant? 88

| Q9.1 - If you use voice interfaces, what aspects of your speech patterns change when using a voice assistant? | Percentage | Count |
| --- | --- | --- |
| Enunciation | 61% | 54 |
| Volume | 52% | 46 |
| Sentence complexity | 52% | 46 |
| Number of syllables in word choices | 18% | 16 |
| Length of sentences | 38% | 33 |
| Use of conjunctions | 16% | 14 |
| Pauses and flow | 51% | 45 |

If you use AI-powered chats, what aspects of your sentence structure changes when engaging with the AI? 81

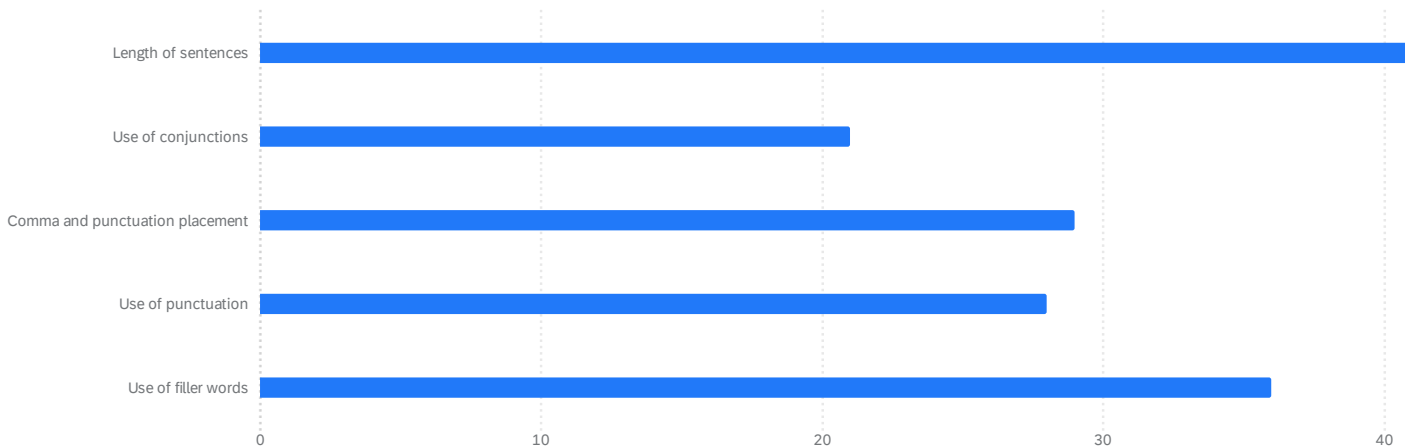

If you use AI-powered chats, what aspects of your sentence structure changes when engaging with the AI? 81

| Q9.2 - If you use AI-powered chats, what aspects of your sentence structure changes when engaging with the AI? | Percentage | Count |
| --- | --- | --- |
| Length of sentences | 52% | 42 |
| Use of conjunctions | 26% | 21 |
| Comma and punctuation placement | 36% | 29 |
| Use of punctuation | 35% | 28 |
| Use of filler words | 44% | 36 |
| Sum | 193% | 156 |

# Communication with technology / Sentiment

Responses: 101

**I trust AI to understand me correctly.** 89

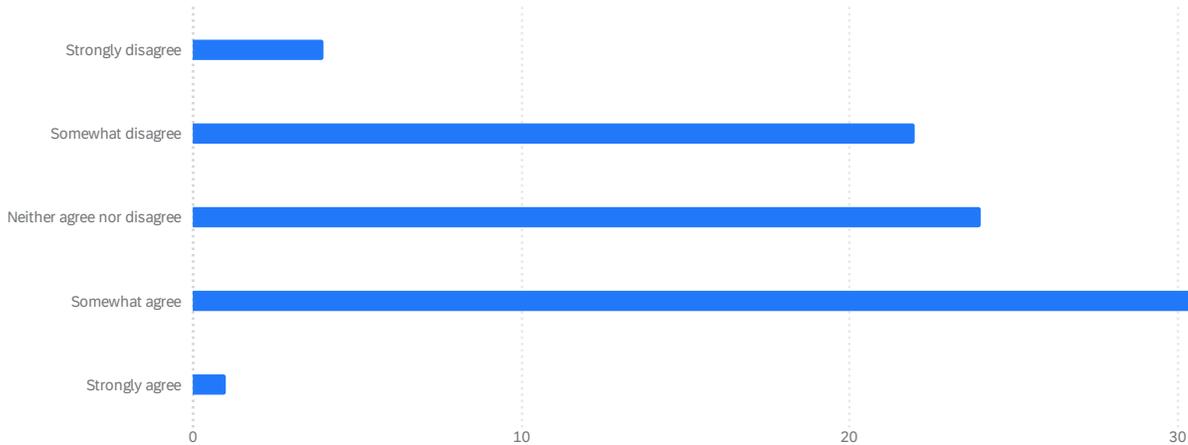

**I trust AI to understand me correctly.** 89

| Q10.2 - I trust AI to understand me correctly. | Percentage | Count |
|---|---|---|
| Strongly disagree | 4% | 4 |
| Somewhat disagree | 25% | 22 |
| Neither agree nor disagree | 27% | 24 |
| Somewhat agree | 43% | 38 |
| Strongly agree | 1% | 1 |
| Sum | 100% | 89 |

**I can converse naturally and casually with AI and I will be understood.** 89

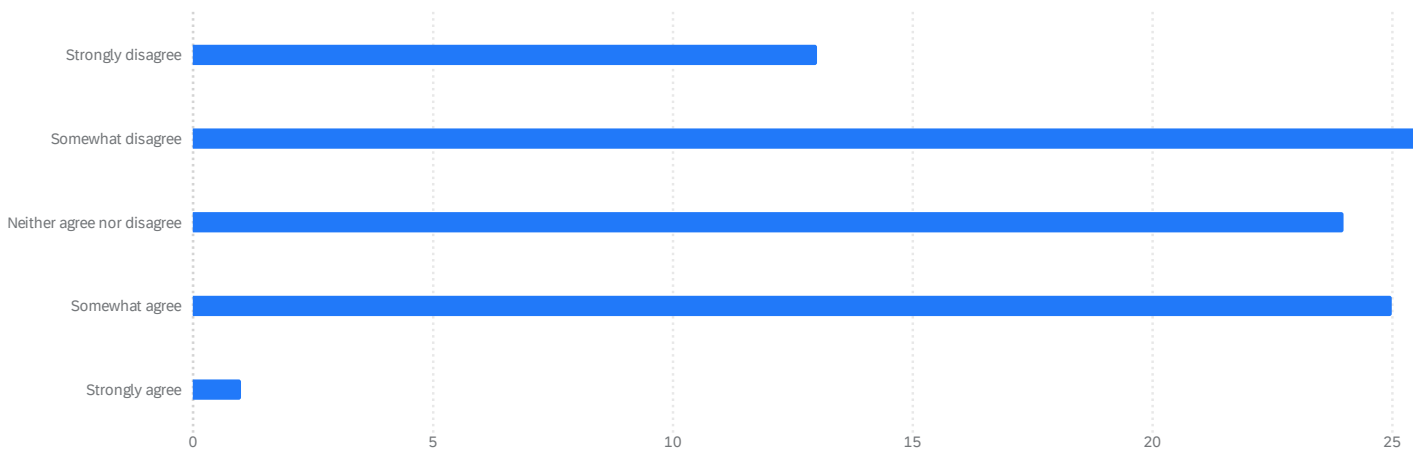

### I can converse naturally and casually with AI and I will be understood. 89 ⓘ

| Q10.3 - I can converse naturally and casually with AI and I will be understood. | Percentage | Count |
|---|---|---|
| Strongly disagree | 15% | 13 |
| Somewhat disagree | 29% | 26 |
| Neither agree nor disagree | 27% | 24 |
| Somewhat agree | 28% | 25 |
| Strongly agree | 1% | 1 |
| Sum | 100% | 89 |

### AI works well. 89 ⓘ

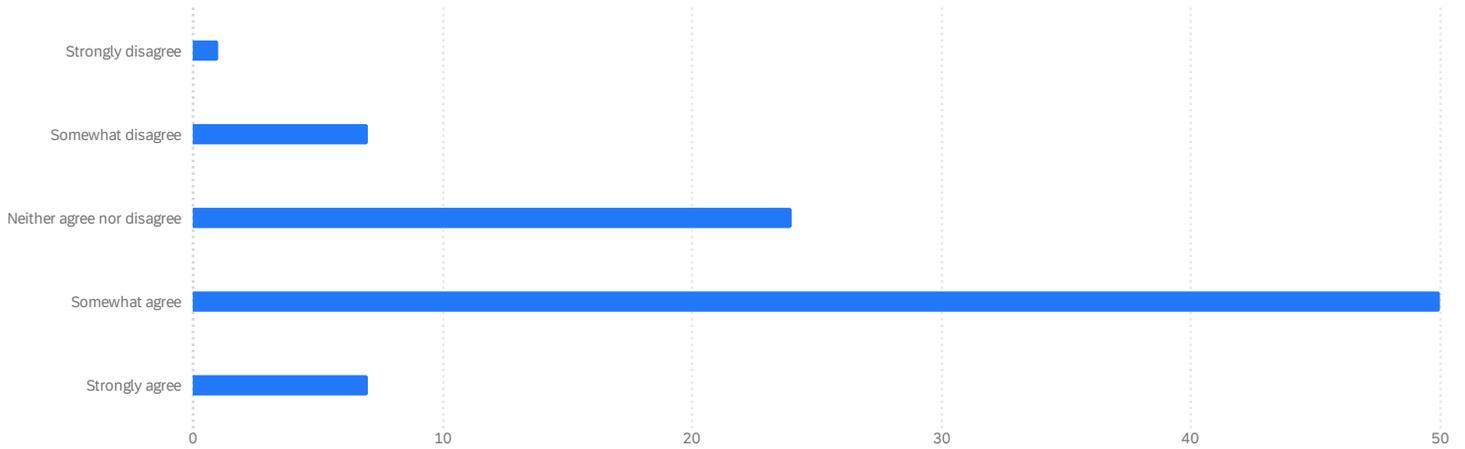

### AI works well. 89 ⓘ

| Q10.4 - AI works well. | Percentage | Count |
|---|---|---|
| Strongly disagree | 1% | 1 |
| Somewhat disagree | 8% | 7 |
| Neither agree nor disagree | 27% | 24 |
| Somewhat agree | 56% | 50 |
| Strongly agree | 8% | 7 |
| Sum | 100% | 89 |

I rarely have to repeat myself when conversing with an AI. 89

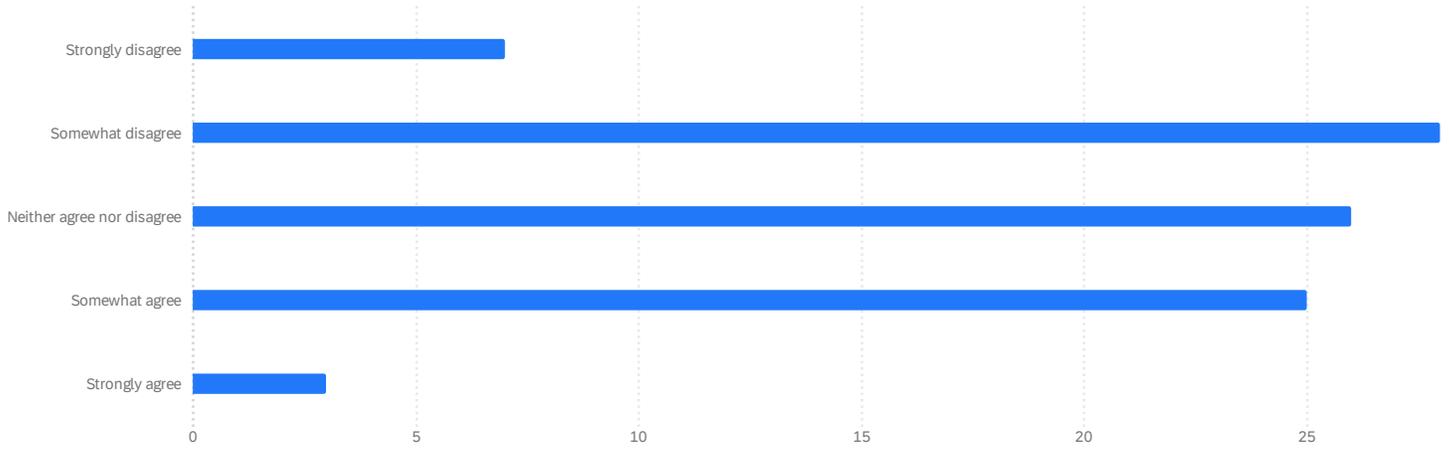

I rarely have to repeat myself when conversing with an AI. 89

| Q10.5 - I rarely have to repeat myself when conversing with an AI. | Percentage | Count |
| --- | --- | --- |
| Strongly disagree | 8% | 7 |
| Somewhat disagree | 31% | 28 |
| Neither agree nor disagree | 29% | 26 |
| Somewhat agree | 28% | 25 |
| Strongly agree | 3% | 3 |
| Sum | 100% | 89 |

I phrase things when conversing with an AI in the same way I would with a friend. 89

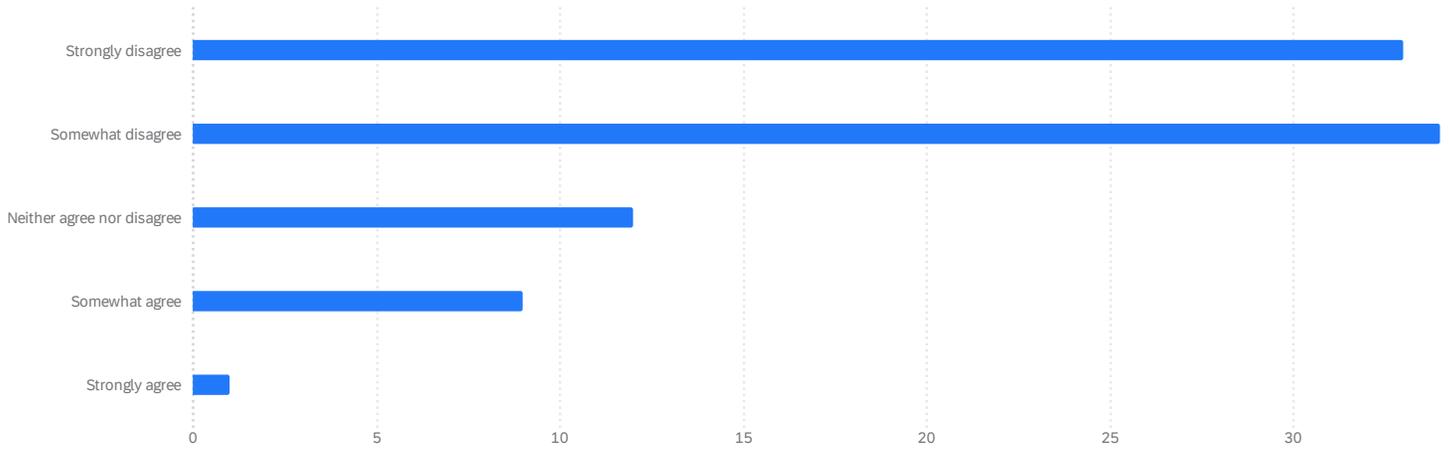

## I phrase things when conversing with an AI in the same way I would with a friend. 89 ⓘ

| Q10.6 - I phrase things when conversing with an AI in the same way I would with a friend. | Percentage | Count |
|---|---|---|
| Strongly disagree | 37% | 33 |
| Somewhat disagree | 38% | 34 |
| Neither agree nor disagree | 13% | 12 |
| Somewhat agree | 10% | 9 |
| Strongly agree | 1% | 1 |
| Sum | 100% | 89 |

## I pay attention to my sentence structure more when conversing with an AI than with a friend. 89 ⓘ

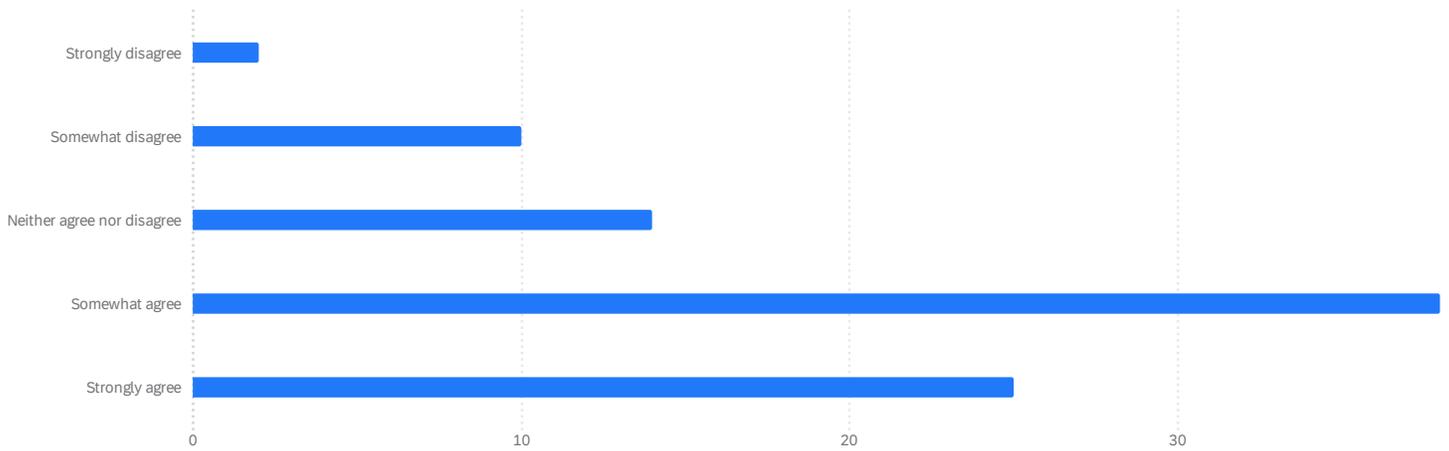

## I pay attention to my sentence structure more when conversing with an AI than with a friend. 89 ⓘ

| Q10.7 - I pay attention to my sentence structure more when conversing with an AI than with a friend. | Percentage | Count |
|---|---|---|
| Strongly disagree | 2% | 2 |
| Somewhat disagree | 11% | 10 |
| Neither agree nor disagree | 16% | 14 |
| Somewhat agree | 43% | 38 |
| Strongly agree | 28% | 25 |
| Sum | 100% | 89 |

I can trust the responses I get from an AI. 89 ⓘ

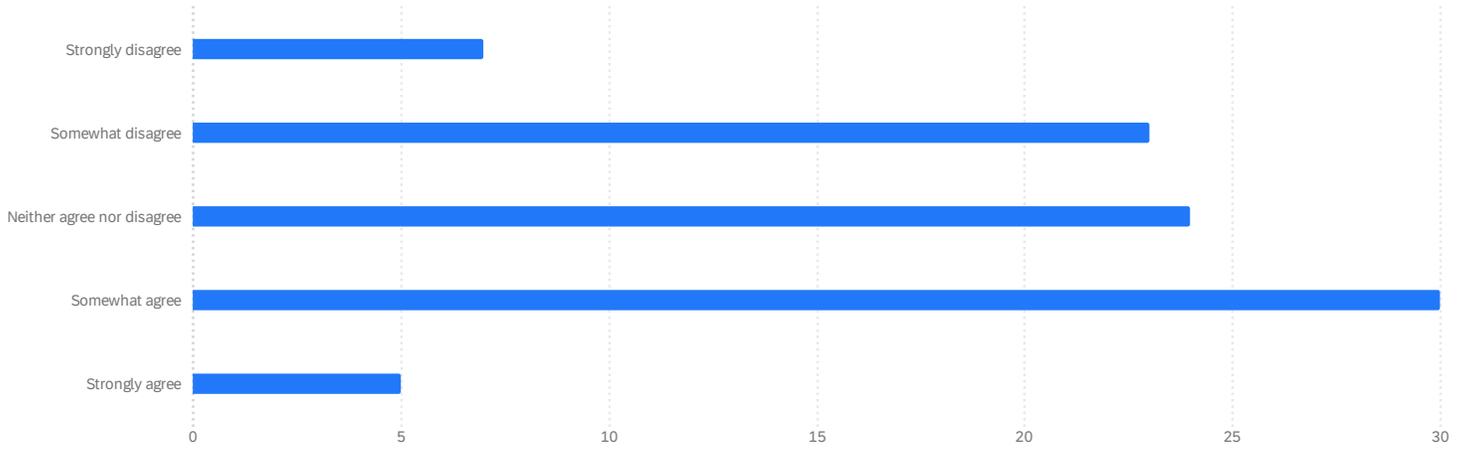

I can trust the responses I get from an AI. 89 ⓘ

| Q10.8 - I can trust the responses I get from an AI. | Percentage | Count |
| --- | --- | --- |
| Strongly disagree | 8% | 7 |
| Somewhat disagree | 26% | 23 |
| Neither agree nor disagree | 27% | 24 |
| Somewhat agree | 34% | 30 |
| Strongly agree | 6% | 5 |
| Sum | 100% | 89 |

Most reasons why I would contact customer service can be resolved by an AI. 89 ⓘ

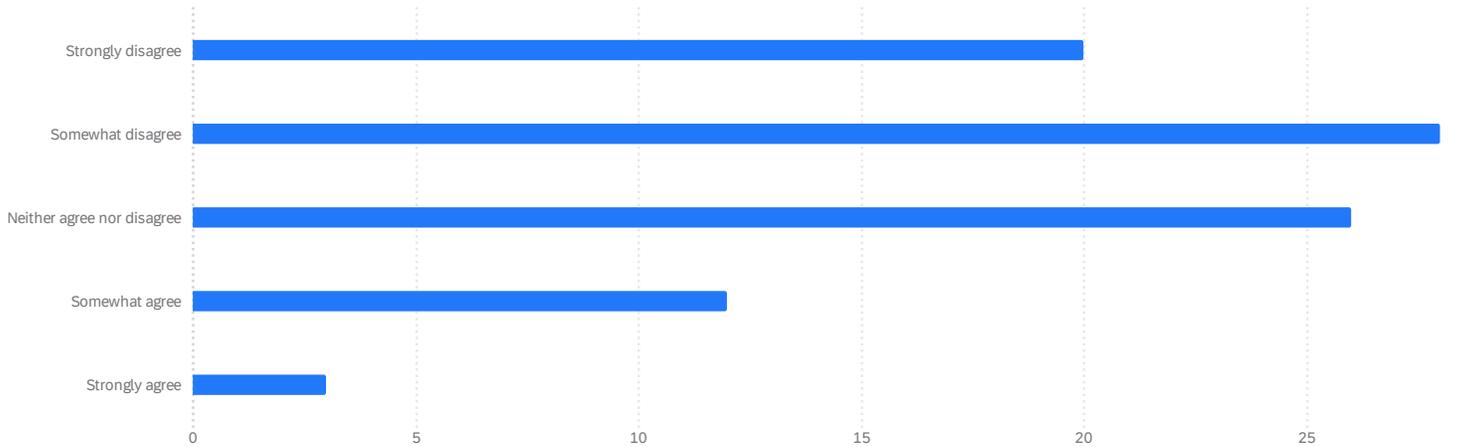

### Most reasons why I would contact customer service can be resolved by an AI. 89

| Q10.9 - Most reasons why I would contact customer service can be resolved by an AI. | Percentage | Count |
|---|---|---|
| Strongly disagree | 22% | 20 |
| Somewhat disagree | 31% | 28 |
| Neither agree nor disagree | 29% | 26 |
| Somewhat agree | 13% | 12 |
| Strongly agree | 3% | 3 |
| Sum | 100% | 89 |

### I change how I phrase things to AIs to try to help it understand me better. 89

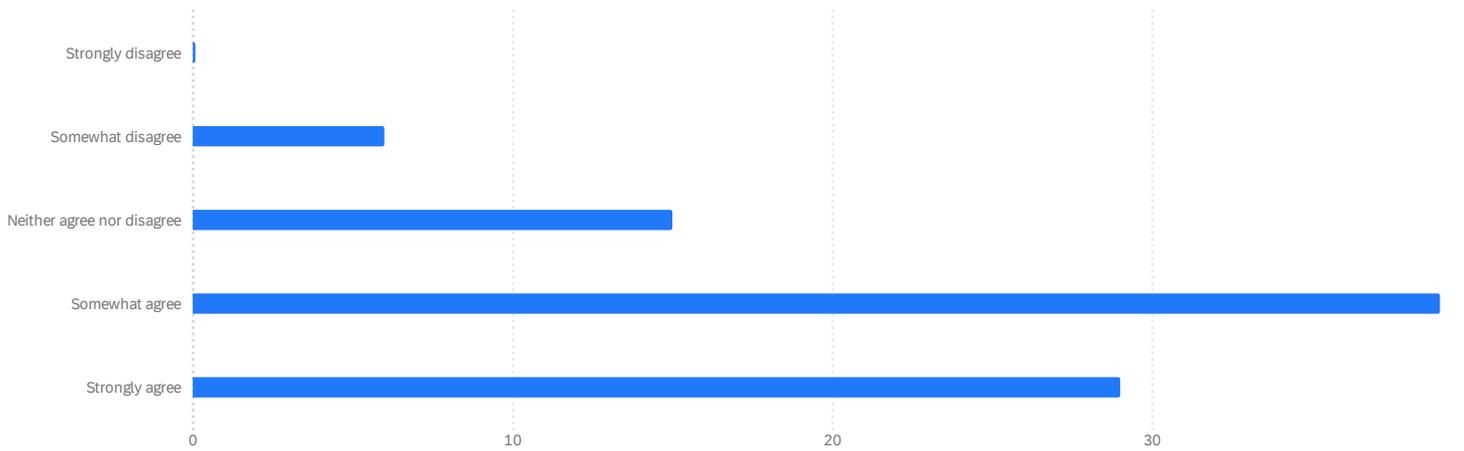

### I change how I phrase things to AIs to try to help it understand me better. 89

| Q10.10 - I change how I phrase things to AIs to try to help it understand me better. | Percentage | Count |
|---|---|---|
| Strongly disagree | 0% | 0 |
| Somewhat disagree | 7% | 6 |
| Neither agree nor disagree | 17% | 15 |
| Somewhat agree | 44% | 39 |
| Strongly agree | 33% | 29 |
| Sum | 100% | 89 |

I think AI has limitations when it comes to understanding the way I speak or write. 89

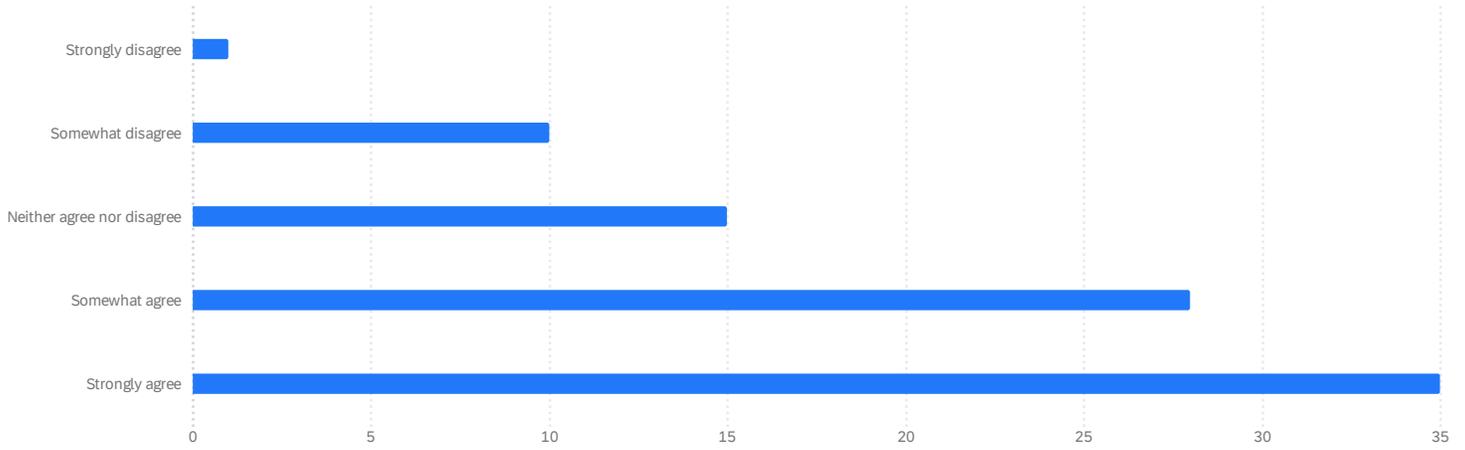

I think AI has limitations when it comes to understanding the way I speak or write. 89

| Q10.11 - I think AI has limitations when it comes to understanding the way I speak or write. | Percentage | Count |
| --- | --- | --- |
| Strongly disagree | 1% | 1 |
| Somewhat disagree | 11% | 10 |
| Neither agree nor disagree | 17% | 15 |
| Somewhat agree | 31% | 28 |
| Strongly agree | 39% | 35 |
| Sum | 100% | 89 |

I have to think more systematically to communicate with an AI. 89

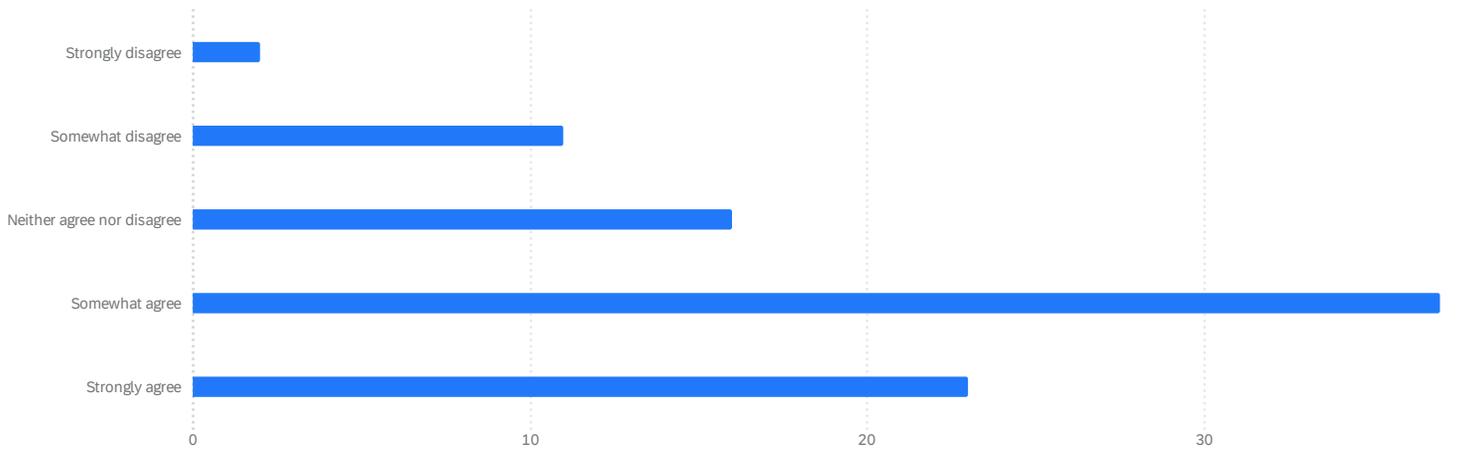

## I have to think more systematically to communicate with an AI. 89

| Q10.12 - I have to think more systematically to communicate with an AI. | Percentage | Count |
| --- | --- | --- |
| Strongly disagree | 2% | 2 |
| Somewhat disagree | 12% | 11 |
| Neither agree nor disagree | 18% | 16 |
| Somewhat agree | 42% | 37 |
| Strongly agree | 26% | 23 |
| Sum | 100% | 89 |

## I must be literal when communicating with an AI for it to understand me. 89

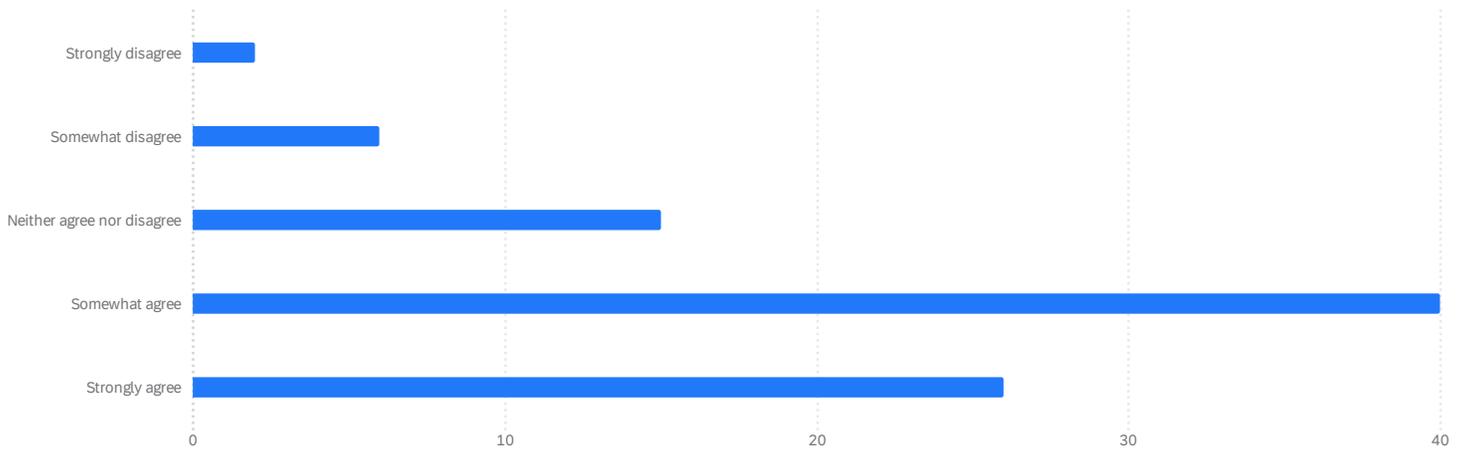

## I must be literal when communicating with an AI for it to understand me. 89

| Q10.13 - I must be literal when communicating with an AI for it to understand me. | Percentage | Count |
| --- | --- | --- |
| Strongly disagree | 2% | 2 |
| Somewhat disagree | 7% | 6 |
| Neither agree nor disagree | 17% | 15 |
| Somewhat agree | 45% | 40 |
| Strongly agree | 29% | 26 |
| Sum | 100% | 89 |

Interacting with an AI feels natural. 89

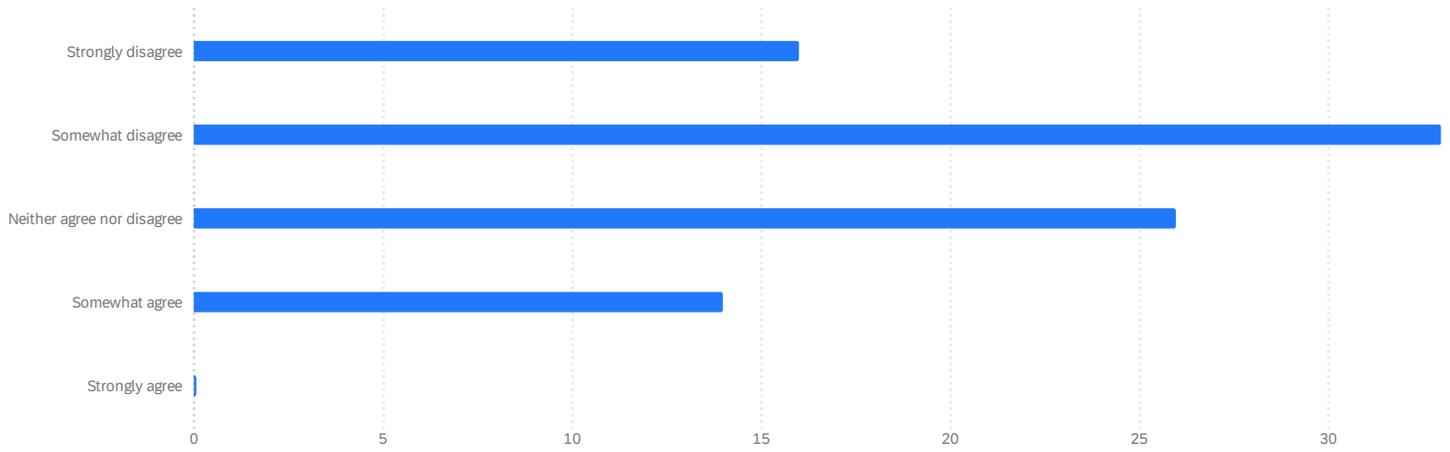

Interacting with an AI feels natural. 89

| Q10.14 - Interacting with an AI feels natural. | Percentage | Count |
| --- | --- | --- |
| Strongly disagree | 18% | 16 |
| Somewhat disagree | 37% | 33 |
| Neither agree nor disagree | 29% | 26 |
| Somewhat agree | 16% | 14 |
| Strongly agree | 0% | 0 |
| Sum | 100% | 89 |